\documentclass[%
superscriptaddress,
 amsmath,amssymb,
 aps,
 pra,
floatfix,
twocolumn
]{revtex4-2}
\usepackage[utf8]{inputenc}
\usepackage[T1]{fontenc}
\usepackage{babel}
\usepackage{graphicx}
\usepackage{amsfonts}
\usepackage{amssymb}
\usepackage{amsmath}
\usepackage{amsthm}
\usepackage{mathtools}
\usepackage{physics}
\usepackage[normalem]{ulem}
\usepackage{blindtext}

\DeclareMathOperator{\diag}{diag}

\DeclareMathOperator{\re}{re}
\DeclareMathOperator{\im}{im}

\usepackage{braket}
\renewcommand\bra[1]{{\langle{#1}|}}
\renewcommand\ket[1]{{|{#1}\rangle}}
\usepackage{enumerate}
\usepackage{dsfont}
\usepackage{bm}
\usepackage{xcolor}

\newtheorem{theorem}{Theorem}
\newtheorem{proposition}[theorem]{Proposition}
\newtheorem{corollary}[theorem]{Corollary}
\theoremstyle{definition}

\usepackage{natbib}
\setcitestyle{square,numbers}
\usepackage[colorlinks]{hyperref}

\begin{document}

\author{Tomasz Linowski}
\affiliation{International Centre for Theory of Quantum Technologies, University of Gdansk, 80-308 Gda{\'n}sk, Poland}
\email[Corresponding author: ]{linowski@cft.edu.pl}

\author{{\L}ukasz Rudnicki}
\affiliation{International Centre for Theory of Quantum Technologies, University of Gdansk, 80-308 Gda{\'n}sk, Poland}
\affiliation{Center for Theoretical Physics, Polish Academy of Sciences, 02-668 Warszawa, Poland}

\title{Reduced state of the field and classicality of quantum Gaussian evolution}

\date{\today}

\begin{abstract}
The notion of classicality of quantum evolution of light is an object of both conceptual and practical importance. The main goal of this work is to derive the exact conditions for the classicality of quantum Gaussian evolution, i.e. the evolution of Gaussian states of light and their convex combinations, a model which is of great significance in quantum optics and information. Several examples, ranging from Gaussian thermal operations to entanglement-maximizing dissipative engineering, are discussed. Our results are obtained using the recently introduced mesoscopic theory of the reduced state of the field, which was originally devised as as a description of macroscopic quantum fields. Here, to make the framework suitable for our goal, we redevelop it as a tool for probing classicality, which constitutes our second main contribution.
\end{abstract}

\maketitle

\section{Introduction}
Classicality of light has been a subject of an ongoing debate at least since Einstein's work on photoelectric effect \cite{photoelectric_effect_Einstein_1905} and the discovery of wave-particle duality \cite{wave_particle_duality_Bohr_1928}. While it is generally believed that, e.g., Glauber's coherent states \cite{P_representation_Glauber} are more classical than pure Fock states with the same mean particle number \cite{coherent_states_review_Martin-Dussaud_2021}, there exists no widely accepted criterion for classicality of multi-photon states of light. Even classicality of a single photon continues to be vividly discussed \cite{one_photon_classicality_Markiewicz_2019,one_photon_classicality_Das_2022}.

Similar considerations concern the time evolution of the electromagnetic field. Quantum particles evolve under the von Neumann equation, while classical particles evolve under Liouville's equation. The degree to which the latter approximates the former is quantified by the relation between the energy scales in the system and the Planck's constant. However, the evolution of both the classical and quantum electromagnetic fields is given by the same set of four Maxwell's equations \cite{quantum_optics_book_Lambropoulos_2007,gaussian_optics}. 

In modern quantum optics, the electromagnetic field is typically described by second quantization, with the occupation numbers of photons of a given frequency described by the density operator in an infinitely-dimensional Hilbert space \cite{gaussian_optics,Gaussian_optics_Olivares_2012}. Currently available experimental operations, which describe the time evolution of this density operator, consist primarily of Gaussian operations and measurements \cite{Gaussianity_resource_Albarelli_2018,Gaussianity_resource_Takagi_2018} and are used, e.g., for quantum key distribution and other information processing tasks \cite{QKD_Gaussian_Grosshans_2002,gaussian_information_1,gaussian_information_2}. 

To this day, the classicality of Gaussian evolution and, more broadly, quantum evolution of light and Gaussian wavepackets, was investigated via a number of methods, relying on phase-space and the Winger distribution, hybridization of quantum and classical theories, and path integrals \cite{Gaussian_classicality_Wigner_Littlejohn_1986,Gaussian_classicality_hybrid_Huber_1988,Gaussian_classicality_hybrid_Huber_1989,Gaussian_classicality_path_integrals_2001}. Besides its conceptual significance, identifying classicality of evolution of light is important in practice, since classical description is typically much simpler than quantum theory \cite{Gaussian_classicality_useful_Javanainen_2013,Gaussian_classicality_useful_Kong_2016,Gaussian_classicality_useful_Graefe_2018}.

In this article, we approach the problem from the point of view of the reduced state of the field (RSF) \cite{alicki_reduced_state}, a recent mesoscopic theory \cite{mesoscopic_theory_Grmela_2015} for many-particle bosonic systems. Relying solely on the first two moments of the mode creation and annihilation operators, the description reduces the infinitely-dimensional density operator of the $N$ mode field to an $N$-dimensional matrix defining the aforementioned RSF. Originally, RSF was designed to describe the potential quantum features of macroscopic fields of a single particle type, including light fields. In particular, the formalism was successfully applied to polarization optics, bridging the Mueller and Jones calculi, as well as to shock wave generation \cite{alicki_reduced_state}. 

Here, we employ RSF to isolate the classical subclass of quantum Gaussian evolution, defined as evolution preserving the set of Gaussian states and their convex combinations. We do this in two steps. First, we investigate the formalism itself to show that, complementarily to its original goal, RSF also captures the classical aspects of quantum fields. In particular, we prove that RSF contains limited information about bipartite entanglement, if any, and that the entropic description in terms of RSF closely resembles that given by the semi-classical Wehrl entropy \cite{wehrl_entropy,wehrl_interpretation}. In this way, we establish RSF as a tool for studies of classicality within quantum mechanics.

Second, we compare quantum Gaussian evolution with the time evolution model built into the RSF framework, deriving in this fashion the explicit subset of Gaussian evolution which is classical with respect to the RSF toolbox. The classicality of the obtained evolution is intuitive, as it consists exclusively of passive transformations, which correspond to experimental operations that can be successfully understood by treating light as a classical wave, such as beam-splitters and phase-shifters. On the contrary, evolution employing quantum squeezing does not fit in the RSF framework.

This work is organized as follows. In Section \ref{sec:Gaussian_preliminaries}, we briefly summarize the main subject of our work -- quantum Gaussian evolution. Over the course of the next three sections, we introduce the formalism of RSF and investigate its various aspects with respect to their classicality: correlations in Section \ref{sec:RSF}, entropy in Section \ref{sec:entropy} and time evolution in Section \ref{sec:reduced_kinetic_equations}. Finally, in Section \ref{sec:evolution}, we use the RSF formalism to derive the classical subset of the quantum Gaussian evolution family. We conclude in Section \ref{sec:conclusion}.

\paragraph*{Notational remark} In this work, we employ three different formalisms: the standard, density operator picture, the symplectic picture and the RSF framework. For clarity, we use different notation for operators in each of these pictures. Operators associated with the the standard picture are denoted by hats, e.g. $\hat{\rho}$. Operators associated with the symplectic picture are denoted by capital letters (with no hats), e.g. $V$. Operators associated with the RSF framework are denoted by small letters (also with no hats), e.g. $r$.

\section{Quantum Gaussian evolution of light}
\label{sec:Gaussian_preliminaries}
We begin by introducing the main subject of our considerations: quantum Gaussian evolution. To this end, we also briefly summarize the notions of Gaussian states and symplectic picture, which will serve as important tools in the derivation of our findings.

\subsection{Gaussian states and symplectic picture}
We consider an $N$-mode Hilbert space spanned by the set of $N$ pairs of mode quadratures collected in the vector
\begin{align} \label{xi}
    \hat{\vec{\xi}} \coloneqq \big(\hat{x}_1, \hat{p}_1, \ldots, \hat{x}_N, \hat{p}_N\big)^T,
\end{align}
where $\hat{x}_k$ and $\hat{p}_k$ fulfill the canonical commutation relations:
\begin{align} \label{eq:canonical_commutation_relations_xi}
    [\hat{x}_k,\hat{p}_{k'}] = i \delta_{kk'}, 
    \qquad [\hat{x}_k,\hat{x}_{k'}] = [\hat{p}_k,\hat{p}_{k'}] = 0
\end{align}
where we set $\hbar=1$. Since the mode quadratures form a basis of operators acting on the $N$-mode Hilbert space, the state of the system is fully described by the complete collection of correlation functions of the form
\begin{align}
\braket{\hat{\xi}_{l_1}\ldots\hat{\xi}_{l_n}}
    \coloneqq\Tr\big[\hat{\rho}\,\hat{\xi}_{l_1}\ldots\hat{\xi}_{l_n}\big].
\end{align}
In the case of Gaussian states, defined as states with normal (Gaussian) characteristic functions and quasiprobability distributions \cite{two-mode_gaussian_etc_proper_norm,cv_systems_gaussian_states,gaussian_optics}, the complete information about the system is contained within only the one- and two-point correlation functions, i.e. with $n=1,2$ in the equation above. The former are contained in the vector of means
\begin{align}
\ket{\xi}\coloneqq\sum_{k=1}^{2N}\braket{\hat{\xi}_k}\ket{k},
\end{align}
while the latter are encoded in the matrix of second moments
\begin{align} \label{covariance_matrix}
    V\coloneqq\frac{1}{2}\sum_{k,k'=1}^{2N}
        \braket{\big\{\hat{\xi}_k,\hat{\xi}_{k'}\big\}}\ket{k}\bra{k'}.
\end{align}
Often, instead of $V$, one uses the covariance matrix, defined as $V_\mathrm{cov}=V-\ket{\xi}\bra{\xi}$. 

Any valid covariance matrix has to fulfill the Heisenberg uncertainty principle:
\begin{align} \label{uncertainty_principle}
\sqrt{\braket{\hat{x}_k^2}-\braket{\hat{x}_k}^2}
	\sqrt{\braket{\hat{p}_k^2}-\braket{\hat{p}_k}^2}
	\geqslant \frac{1}{2},
\end{align}
where $k\in\{1,\ldots,N\}$, equivalent to \cite{two-mode_gaussian_etc_proper_norm}
\begin{align} \label{eq:Heisenberg_uncertainty_principle}
    V_\mathrm{cov}-\frac{i}{2}J \geqslant 0.
\end{align}
Here, $J$ is the \emph{symplectic form}, defined as
\begin{align} \label{symplectic_form}
    J \coloneqq -i\sum_{k,k'=1}^{2N} \big[\hat{\xi}_k,\hat{\xi}_{k'}\big]\ket{k}\bra{k'},
\end{align}
and explicitly equal to
\begin{align} \label{symplectic_form_explicit}
    J = \bigoplus_{k=1}^N J_2, 
        \quad J_2 \coloneqq
        \begin{bmatrix}
        0&1\\
        -1&0
        \end{bmatrix}.
\end{align}
The symplectic form defines the symplectic group $Sp(2N,\mathbb{R})$ consisting of matrices $S$ of size $2N \times 2N$, such that $SJS^T=J$. 

As a matter of fact, the pair $(V,\ket{\xi})$ contains the same information as $(V_\mathrm{cov},\ket{\xi})$, and both in the same way define the \emph{symplectic picture} of quantum states (sometimes referred to as covariance matrix picture), which is a convenient description of the first two moments of the system, particularly in the case of Gaussian states and dynamics. Here, we employ the pair $(V,\ket{\xi})$, since, as we will see in the next section, it is by construction closer to the reduced state of the field than $(V_\mathrm{cov},\ket{\xi})$.

\subsection{Quantum Gaussian time evolution}
The time evolution of quantum open systems is typically modeled by the \emph{Gorini-Kossakowski-Lindblad-Sudarshan (GKLS) equation} (also known as the Lindblad equation) \cite{GKS_original,lindblad_original,lindblad_proof_mathematical}, which in the diagonalized form reads:
\begin{align} \label{eq:GKLS_diagonal}
    \frac{d}{dt}\hat{\rho}=&
        -i\big[\hat{H},\hat{\rho}\big]
        +\sum_{k}\left(\hat{L}_k\hat{\rho}\hat{L}_{k}^\dag
        -\frac{1}{2}\big\{\hat{L}_{k}^\dag\hat{L}_{k},\hat{\rho}\big\}\right),
\end{align}
where $\hat{H}$ denotes the system Hamiltonian and $\hat{L}_k$ are the Lindblad operators.

One of the main sources of motivation for studying Gaussian states is that, due to technical limitations, in practice we are often restricted to Hamiltonians that are polynomials of at most second degree in mode quadratures:
\begin{align} \label{eq:hamiltonian_quadratic}
    \hat{H} = \frac{1}{2}\hat{\vec{\xi}}^T G \hat{\vec{\xi}},
\end{align}
where $G$ is a $2N\times 2N$, real, symmetric matrix. The structure-preserving evolution of Gaussian states is driven by precisely such Hamiltonians. 

Similarly, to preserve Gaussianity of an initial state along the course of time evolution, the Lindblad operators need to be linear in mode quadratures \cite{WolfHolevo}:
\begin{align} \label{eq:Lindbladians_linear}
    \hat{L}_k=\vec{c}_k \cdot \hat{\vec{\xi}},\quad \vec{c}_k\in\mathbb{C}^{2N},
\end{align}
so that the resulting dissipator is a polynomial of second degree in mode quadratures, like the Hamiltonian.

However, the same experimental tools that let one create and manipulate Gaussian states can be used to create and manipulate convex combinations of Gaussian states. In fact, according to recent theories of non-Gaussianity \cite{Gaussianity_resource_Albarelli_2018,Gaussianity_resource_Takagi_2018}, from the point of view of useful non-Gaussianity there is no difference between Gaussian states and their convex combinations. Only states that cannot be written as a convex combination of Gaussian states are genuinely non-Gaussian, or \emph{quantum non-Gaussian} \cite{quantum_Gaussianity_Genoni_2013,quantum_Gaussianity_Hughes_2014,quantum_Gaussianity_Walschaers_2021}. Consequently, states that are either Gaussian or can be written as a convex combination of Gaussian states are called \emph{quantum Gaussian}.

For this reason, in addition to linear Lindblad operators, which preserve the set of Gaussian states, we also consider unitary Lindblad operators
\begin{align} \label{eq:Lindbladians_unitary}
    \hat{L}_k = \sqrt{\gamma_k} \hat{U}_k,
\end{align}
where $\gamma_k\geqslant 0$, $\sum_k \gamma_k = 1$ and $\hat{U}_k$ are unitary operators such that $\hat{U}_k=\exp(i\hat{g}_k)$ with $\hat{g}_k$ being polynomials of at most second degree in the mode quadratures. The dynamics induced by such Lindblad operators does not preserve the set of Gaussian states, but preserve set of quantum Gaussian states \cite{covariance_matrix_beyond_quadratic_linowski_2021}. Dissipators of this form are most well-known for describing random scattering, see \cite{unitary_lindbladians_kossakowski_1972,unitary_lindbladians_Kummerer_1987,unitary_lindbladians_Frigerio_1989}.

Written in the symplectic picture, the evolution given by the Hamiltonian (\ref{eq:hamiltonian_quadratic}) and Lindblad operators (\ref{eq:Lindbladians_linear}, \ref{eq:Lindbladians_unitary}) reads \cite{using_dissipation_1,using_dissipation_2,stabilizability_cv_systems}
\begin{align} \label{eq:covariance_evolution}
\begin{split}
    \frac{d}{dt}V &= F_{G}(V) + F_{L}(V) + F_{U}(V), \\
     \frac{d}{dt}\ket{\xi} &= f_{G}(\ket{\xi}) + f_{L}(\ket{\xi}) + f_{U}(\ket{\xi}).
\end{split}
\end{align}
Here,
\begin{align} \label{eq:F_G}
\begin{split}
    F_{G}(V) & \coloneqq JGV - VGJ, \\ 
    f_G(\ket{\xi}) & \coloneqq JG\ket{\xi},
\end{split}
\end{align}
are the Hamiltonian terms, while \footnote{We note the Hamiltonian term and the ``linear'' dissipative term are typically combined in a single term given by $A=J(G+I_C)$, so that the evolution reads $dV/dt = AV+VA^T +J R_C J^T$. In this work, however, we study the Hamiltonian and the dissipative dynamics separately.}
\begin{align} \label{eq:F_L}
\begin{split}
    F_{L}(V) & \coloneqq J I_C V + V I_C J + J R_C J^T, \\  
    f_L(\ket{\xi}) & \coloneqq J I_C \ket{\xi},
\end{split}
\end{align}
with $R_C\equiv\re C^\dag C$, $I_C\equiv\im C^\dag C$ and $C_{kl}\coloneqq (\vec{c}_k)_{l}$ stem from linear Lindblad operators (\ref{eq:Lindbladians_linear}). We remark that the Gaussianity-preserving time evolution given by these functions is known to have exact solutions \cite{Gaussian_solvable_Hu_1992,Gaussian_solvable_Karrlein_1997,Gaussian_solvable_Zhang_2012} and is well-studied using Green functions \cite{Gaussian_Green_Wang_2014,Gaussian_Green_Dhar_2014} and, in particular, the symplectic picture \cite{using_dissipation_1,using_dissipation_2,stabilizability_cv_systems}.

On the other hand, the final terms 
\begin{align} \label{eq:F_U}
\begin{split}
     F_{U}(V)&\coloneqq \sum_j \gamma_j \left(K_j V K_j^T - V\right),\\
     f_{U}(\ket{\xi})&\coloneqq \sum_j \gamma_j \left(K_j \ket{\xi} - \ket{\xi}\right),\\
\end{split}
\end{align}
where $K_j$ are symplectic, stem from the unitary Lindblad operators (\ref{eq:Lindbladians_unitary}) and represent a relatively novel type of dynamics that preserves only the set of quantum Gaussian states \cite{covariance_matrix_beyond_quadratic_linowski_2021}.

Eq. (\ref{eq:covariance_evolution}) defines the \emph{quantum Gaussian evolution}. The ultimate goal of our article is to identify the subclass of semi-classical evolution consistent with this equation. Before we can do that, however, we need to develop the necessary tools to achieve this goal, namely, the framework of the reduced state of the field (RSF).

\section{Reduced state of the field as a classical description of bosonic fields} 
\label{sec:RSF}
In this section, we summarize the relevant information about RSF and simultaneously investigate it to show that it provides a semi-classical description for bosonic many-particle fields, thus constituting a viable tool for our main goal.

\subsection{Reduced state of the field (RSF)}
The main idea behind RSF was to describe many-particle, or macroscopic, quantum fields. In such a case, instead of using the mode quadratures, it is often more convenient to use the annihilation and creation operators 
\begin{align} \label{eq:creation_annihilation_operators}
    \hat{a}_k\coloneqq
    \frac{1}{\sqrt{2}}\left(\hat{x}_k+i\hat{p}_k\right), \quad
    \hat{a}_{k}^\dag =
    \frac{1}{\sqrt{2}}\left(\hat{x}_k-i\hat{p}_k\right),
\end{align}
with the canonical commutation relations (\ref{eq:canonical_commutation_relations_xi}) now reading
\begin{align} \label{eq:canonical_commutation_relations}
    \big[\hat{a}_k,\hat{a}_{k'}^\dag\big]=\delta_{kk'},
    \quad \big[\hat{a}_k,\hat{a}_{k'}\big]
    =\big[\hat{a}_k^\dag,\hat{a}_{k'}^\dag\big]=0.
\end{align}
An arbitrary $n$-particle state in the many-body Hilbert space can be then constructed by acting on the vacuum state with $n$ appropriate creation operators.

In the case of macroscopic fields, typically modeled as non-interacting fields with dynamics governed by field equations linear in creation and annihilation operators with possible external coherent sources, the fundamental observables are either \emph{additive}, like energy \cite{alicki_reduced_state}:
\begin{align} \label{eq:observable_additive}
    \hat{O} = \sum_{k,k'=1}^N o_{kk'} \hat{a}_k^\dag \hat{a}_{k'}
\end{align}
or \emph{linear}, like momentum: 
\begin{align} \label{eq:observable_linear}
    \hat{\sigma} = \sum_{k=1}^N \left( \sigma_{k}^* \hat{a}_k + \sigma_k \hat{a}_k^\dag \right).
\end{align}
One can easily check that the expectation values of such observables can be equivalently rewritten as
\begin{align} \label{eq:observables_translation}
\begin{split}
    \Tr \hat{\rho} \, \hat{O} & = \tr r o, \qquad
    \Tr \hat{\rho} \, \hat{\sigma} = \braket{\sigma|\alpha} + \braket{\alpha|\sigma},
\end{split}
\end{align}
where 
\begin{align} \label{eq:single_particle_density_matrix_definition}
    {r}\coloneqq\sum_{k,k'=1}^N
        \Tr\big[\hat{\rho}\,\hat{a}_{k'}^\dag\hat{a}_k\big]
        \ket{k}\bra{k'}
\end{align}
defines the \emph{single-particle density matrix},
\begin{align} \label{eq:averaged_field_defition}
    \ket{\alpha}\coloneqq\sum_{k=1}^{N}
        \Tr\big[\hat{\rho}\,\hat{a}_k\big]
        \ket{k}
\end{align}
defines the \emph{averaged field}, while 
\begin{align}
    o = \sum_{k,k'=1}^N o_{kk'} \ket{k} \bra{k'}, 
        \qquad \ket{\sigma} = \sum_{k=1}^N \sigma_{k} \ket{k}
\end{align}
are the \emph{reduced observables} corresponding to $\hat{O}$ and $\hat{\sigma}$.

The single-particle density matrix contains information about mode occupation and coherence in the state. In particular, its diagonal elements equal the mean particle numbers: ${r}_{kk} = \braket{\hat{a}_{k}^\dag\hat{a}_{k}} = \braket{\hat{n}_{k}}$ and, consequently, the matrix is normalized to the mean total particle number: $\tr r = \sum_{k=1}^N\braket{\hat{n}_{k}} \equiv \braket{\hat{n}}$. Note that, by construction, the single-particle density matrix is non-negative. 

The averaged field, on the other hand, contains information about local phases of the field. For example, in the case of a pure Fock state, the phase is undefined, yielding no such information, with the opposite situation in the case of a pure coherent state.

Together, the single particle density matrix and the averaged field constitute the \emph{reduced state of the field (RSF)} associated with the density operator $\hat{\rho}$ \cite{alicki_reduced_state}. A major advantage of the RSF formalism is that it is, to a large degree, self-contained, in the sense that it allows for study of a variety of phenomena without having to refer to other frameworks. In particular, it comes equipped with its own definition of entropy and a time evolution model, both of which are investigated by us in the subsequent sections.

In the case of additive and linear observables (\ref{eq:observable_additive}, \ref{eq:observable_linear}), the RSF description is complete. In the case where the observables of interest are more general, RSF describes a subset of degrees of freedom of the system. We now proceeed to give a physical interpretation for the degrees of freedom contained within RSF.

\subsection{Physical meaning of correlations within the RSF framework}
To see what physical information is associated with the degrees of freedom contained within RSF, we begin by observing that RSF is related to the symplectic picture of quantum states via
\begin{align} \label{eq:reduced_fields_relation_to_covariance_matrix}
\begin{split}
    {r}&=\mathcal{R}V\mathcal{R}^\dag-\frac{{1}}{2}\mathds{1}_N,
        \quad \ket{\alpha}=\mathcal{R}\ket{\xi},
\end{split}
\end{align}
where we use $\mathds{1}_N$ to denote the identity matrix of size $N\times N$ and
\begin{align} \label{eq:transfer_matrix}  
    \mathcal{R}\coloneqq\frac{1}{\sqrt{2}}\sum_{k=1}^N\ket{k}\big[\bra{2k-1}+i\bra{2k}\big]
\end{align}
defines the \emph{reduction matrix}. The Heisenberg uncertainty principle (\ref{eq:Heisenberg_uncertainty_principle}) translates to non-negativity of the \emph{correlation matrix}:
\begin{align} \label{eq:Heisenberg_uncertainty_principle_RSF}
\begin{split}
    r_\alpha \coloneqq r - \ket{\alpha}\bra{\alpha} \geqslant 0,
\end{split}
\end{align}
which was defined already in \cite{alicki_reduced_state}. The relations (\ref{eq:reduced_fields_relation_to_covariance_matrix}, \ref{eq:Heisenberg_uncertainty_principle_RSF}) are derived by us in Appendix \ref{app:reduction_map}.

The input of the reduction matrix belongs to a $2N$-dimensional space, while the output is only $N$-dimensional. Clearly, then, the reduction matrix cuts some of the information from the symplectic picture. As we will now show, this missing information is relevant for practical scenarios requiring bipartite quantum entanglement. In order for a given entangled state to be useful for any such task, e.g. quantum code encryption or teleportation, it first needs to be \emph{distilled} \cite{quantum_resource_theory_entanglement_HHHH_2009}. 

Crucially, not every entangled state is distillable. A necessary condition for bipartite entanglement distillation is given by the \emph{positive partial transpose (PPT) criterion} \cite{PPT,PPT_cv_systems}, originally stated as a necessary condition for separability. Adopted to the language of distillable entanglement, the PPT criterion states that if the partial transposition of the state with respect to a given bipartition is positive semi-definite, then the state does not contain distillable entanglement with respect to this bipartition \cite{quantum_resource_theory_entanglement_HHHH_2009}. 

In the symplectic picture, partial transposition of arbitrary chosen modes is performed by replacing the covariance matrix by
\begin{align}
    V_{\textnormal{cov},\vec{q}} = Q_{\vec{q}}V_{\textnormal{cov}}Q_{\vec{q}},
\end{align}
with
\begin{align}
\begin{split}
    Q_{\vec{q}}=\diag(1,q_1,\ldots,1,q_N),
\end{split}
\end{align}
where $q_k=-1$ for modes that are being transposed and $q_k=1$ otherwise.

From the perspective of distillable entanglement, the PPT criterion for continuous variable systems states that if [cf. eq. (\ref{eq:Heisenberg_uncertainty_principle})]:
\begin{align} \label{eq:PPT}
    V_{\textnormal{cov},\vec{q}}-\frac{i}{2}J \geqslant 0,
\end{align}
the state does not contain distillable entanglement with respect to the bipartion given by $\vec{q}$ \cite{PPT_cv_systems,quantum_resource_theory_entanglement_HHHH_2009}. Therefore, violation of (\ref{eq:PPT}) indicates its presence. Note that if this inequality holds, the state may still contain so-called \emph{bound} entanglement. This type of entanglement is, however, much less useful in practice.

We will now show that in the RSF picture the inequality (\ref{eq:PPT}) is always satisfied. In turn, the formalism contains no information about distillable entanglement. To this end, it is enough to limit our considerations to the correlation matrix (\ref{eq:Heisenberg_uncertainty_principle_RSF}), since the averaged field contains only local information and is therefore irrelevant for entanglement.

The key observation is that among all the covariance matrices corresponding to a given correlation matrix through eq. (\ref{eq:reduced_fields_relation_to_covariance_matrix}), there is one that equals
\begin{align} \label{eq:V_special}
    \bar{V}_{\textnormal{cov}} = r_\alpha \otimes \mathds{1}_2 + \frac{1}{2}\mathds{1}_{2N}.
\end{align}
This can be seen as follows. Firstly, as is easy to compute, $\mathcal{R}( r_\alpha \otimes \mathds{1}_2)\mathcal{R}^\dag= r_\alpha$, from which it immediately follows that the correlation matrix corresponding to $\bar{V}_{\textnormal{cov}}$ is equal to $r_\alpha$. Secondly, $\bar{V}_{\textnormal{cov}}$ fulfills the Heisenberg uncertainty principle (\ref{eq:Heisenberg_uncertainty_principle}), since
\begin{align} \label{eq:V_cov_bar_positive}
    \bar{V}_{\textnormal{cov}}-\frac{i}{2}J 
	= r_\alpha \otimes \mathds{1}_2 +  \frac{1}{2} \left(\mathds{1}_{2N} - i J \right).
\end{align}
By construction, $r_\alpha$ is non-negative, and thus so is the first term on the r.h.s. 
The second term, on the other hand, can be decomposed into $N$ blocks of size $2\times 2$ as $\mathds{1}_{2N} - i J = \bigoplus_{j=1}^N (\mathds{1}_{2} - i J_2)$. It is straightforward to calculate that each of these blocks is non-negative, making the whole matrix non-negative. Therefore, eq. (\ref{eq:V_cov_bar_positive}) is non-negative and $\bar{V}_{\textnormal{cov}}$ is a valid covariance matrix.

Crucially, all variants of partial transposition of $\bar{V}_{\textnormal{cov}}$ satisfy the condition (\ref{eq:PPT}). Indeed, for any $\vec{q}$ we have
\begin{align}
    \bar{V}_{\textnormal{cov},\vec{q}} - \frac{i}{2}J 
	= Q_{\vec{q}}(r_\alpha \otimes \mathds{1}_2) Q_{\vec{q}} 
	+  \frac{1}{2} \left(\mathds{1}_{2N} - i J \right),
\end{align}
where we used the fact that $Q_{\vec{q}}^2=\mathds{1}_{2N}$. As before, the second term is non-negative. However, the same is also true for the first term since, due to unitarity and Hermiticity of $Q_{\vec{q}}$, the eigenvalues of $ Q_{\vec{q}}(r_\alpha \otimes \mathds{1}_2) Q_{\vec{q}}$ are the same as the eigenvalues of $r_\alpha \otimes \mathds{1}_2$.

Consequently, every RSF description corresponds to at least one symplectic description of a system that fulfills the PPT criterion, making this criterion trivial from the point of view of the RSF framework. This leads to the following proposition.

\begin{proposition} \label{th:entanglement}
The RSF framework contains no information about bipartite distillable entanglement.
\end{proposition}

This proposition has special consequences for two-mode Gaussian states, for which the PPT criterion is equivalent to the presence of any form of entanglement, not only distillable entanglement \cite{quantum_resource_theory_entanglement_HHHH_2009}: 

\begin{corollary} \label{th:entanglement_Gaussian}
In the case of two-mode Gaussian states, the RSF framework contains no information about any form of entanglement.
\end{corollary}

We conjecture that these findings hold in general, i.e. RSF contains no information about any type of quantum entanglement in any quantum state. Irrespectively, Proposition \ref{th:entanglement} and Corollary \ref{th:entanglement_Gaussian} show that the ability to describe entanglement within the RSF formalism is severely limited, strongly suggesting the framework to be semi-classical.

\section{Classicality of RSF entropy}
\label{sec:entropy}
The fact that RSF contains limited information about entanglement strongly suggests it is a semi-classical formalism. To further reinforce this interpretation, in this section, we analyze the entropic description in terms of RSF, showing that it is similar to the one given by the semi-classical Wehrl entropy.

\subsection{Reduced entropy}
The standard choice for quantum (information) entropy is given by the von Neumann entropy \cite{geometry_of_quantum_states}
\begin{align} \label{eq:von_Neumann_entropy}
    S_V(\hat{\rho})\coloneqq -\Tr\hat{\rho}\ln\hat{\rho},
\end{align}
where we set $k_B=1$. Because of its information-theoretic origin as a generalization of the Shannon entropy, the von Neumann entropy is most easily interpreted as a measure of uncertainty about the state of the system. The von Neumann entropy is invariant under all unitary transformations and it attains its minimum value -- zero -- for all pure states. 

To describe entropy within the RSF formalism, one needs to find a way to derive a valid entropy measure that only depends on the components of RSF and not on the density operator.

In \cite{alicki_reduced_state}, this was done with the use of the \emph{maximum entropy principle} \cite{maximum_entropy_principle,maximum_entropy_principle_Thurner_2017}. According to this principle, given only a partial knowledge about a physical system, one should assume the highest possible value of entropy consistent with this knowledge. Interpreting entropy as the amount of uncertainty about the state of the system, the maximum entropy principle means simply that one should not presume to be more certain about the system's state than their knowledge lets them.

For example, if one has absolutely no knowledge about which quantum state the system is in, one should assume it to be maximally mixed, i.e. $\hat{\rho}= \hat{\mathds{1}}_\Omega/\Omega$, where $\Omega$ denotes the number of possible orthogonal system states, or equivalently the dimension of the Hilbert space. This is because such density operator is the only one for which all system states are equally probable.

In this case, it is easy to calculate that the von Neumann entropy coincides with the classical Boltzmann entropy
\begin{align} \label{eq:Boltzmann_entropy}
    S_B = S_V\left(\hat{\mathds{1}}_\Omega/\Omega\right) = \log \Omega.
\end{align}
Viewed from the perspective of the maximum entropy principle, the Boltzmann entropy is simply the maximum value of the von Neumann entropy consistent with having no knowledge about the quantum state of the system.

In the RSF formalism, the only information we have about the system is its RSF. Thus, according to the maximum entropy principle, we should assume that the system's von Neumann entropy has the highest value possible for a system with that specific RSF. As was calculated in \cite{alicki_reduced_state}, among all the quantum states with the same RSF $(r,\ket{\alpha})$, the von Neumann entropy is maximal for the thermal-like state
\begin{align} \label{eq:thermal-like_state}
    \hat{\rho}_{r,\ket{\alpha}} = \frac{1}{z} D(\vec{\alpha}) \exp\left(-\sum_{k,k'=1}^N r_{kk'}\hat{a}_{k'}^\dag \hat{a}_k\right) D^\dag(\vec{\alpha}),
\end{align}
where
\begin{align} \label{eq:thermal-like_state_z}
    z =  \Tr \exp\left(- \sum_{k,k'=1}^N r_{kk'}\hat{a}_{k'}^\dag \hat{a}_k\right)
\end{align}
and
\begin{align} \label{eq:displacement operator}
    D(\vec{\alpha}) = \exp\left[\sum_{k=1}^N\left(
        \alpha_{k}^* \hat{a}_k - \alpha_{k} \hat{a}_k^\dag
        \right)\right]
\end{align}
is the (unitary) $N$-mode displacement operator.

Thus, the RSF entropy can be defined as \cite{alicki_reduced_state}
\begin{align} \label{eq:reduced_von_Neumann_entropy}
\begin{split}
    s_v(\hat{\rho})& \coloneqq S_V(\hat{\rho}_{r,\ket{\alpha}}) \\
        & = \tr[(r_\alpha+\mathds{1}_N)\ln(r_\alpha+\mathds{1}_N)-r_\alpha\ln r_\alpha]
\end{split}
\end{align}
with the correlation matrix $r_\alpha$ as in eq. (\ref{eq:Heisenberg_uncertainty_principle_RSF}). In accordance with the maximum entropy principle, such entropy, dubbed \emph{reduced entropy} \cite{alicki_reduced_state}, is simply the maximum value of the von Neumann entropy consistent with having no knowledge about the quantum state of the system except for its RSF.

The reduced entropy satisfies the natural condition $s_v(\hat{\rho})\geqslant 0$, with equality if and only if the correlation matrix is equal to zero, which happens only when the density operator of the field is given by a coherent state. In contrast, the von Neumann entropy vanishes for any pure state. 

While based on sound principles, the reduced entropy lacks a clear physical interpretation. We now proceed to investigate the qualitative and quantitative features of this entropy, showing that is has a semi-classical character akin to the Wehrl entropy.

\subsection{Reduced Wehrl entropy and its classical features}
The Wehrl entropy \cite{wehrl_entropy} is defined as the continuous Shannon entropy of the Husimi Q representation of the quantum state: 
\begin{align} \label{eq:Wehrl_entropy}
    S_W(\hat{\rho}) \coloneqq -\int \frac{d^{2N}\vec{\beta}}{\pi^N} 
        Q(\vec{\beta})\ln Q(\vec{\beta}).
\end{align}
Here, $Q(\vec{\beta})=\bra{\vec{\beta}}\hat\rho\ket{\vec{\beta}}$ is the Husimi Q representation \cite{Q_representation} of the state $\hat\rho$, $\ket{\vec{\beta}}$ is an $N$-mode coherent state and the integration is over the real and imaginary parts of every component of the complex vector $\vec\beta$. 

The Wehrl entropy is typically considered to be a semi-classical approximation to the von Neumann entropy, since it is constructed by replacing the quantum density operator in the definition of the von Neumann entropy by its representation $Q(\vec{\beta})$ in the phase-space \cite{Wehrl_general_entropy_properties_1978,wehrl_interpretation}. The two entropies differ significantly. Unlike the von Neumann entropy, the Wehrl entropy never vanishes, and it attains its minimum value, $N$, only for coherent states \cite{wehrl_minimum}. Furthermore, it is not invariant under all unitary transformations of the state. 

Looking at the reduced entropy (\ref{eq:reduced_von_Neumann_entropy}), we can see that it possesses the same qualities. The fact that it is minimized by coherent states was already discussed. As for invariance under unitary operations, consider, e.g. the transformation $\hat{U}^\dag\hat{a}_k\hat{U} = \cosh\mu\,\hat{a}_k + \sinh\mu\,\hat{a}_k^\dag$ with $\mu\neq 0$. From the definitions (\ref{eq:single_particle_density_matrix_definition}, \ref{eq:averaged_field_defition}), we can calculate that $r$ transforms to $r' = \cosh^2\mu\,r+f(\mu)\neq r$, where $f(\mu)$ depends solely on correlations not included in the RSF formalism. Notably, the reduced entropy of $r'$ differs from that of $r$. Finally, we note that by construction, the reduced entropy provides an upper bound to the von Neumann entropy, another quality shared with the Wehrl entropy.

As seen, the reduced entropy resembles the Wehrl entropy more than the von Neumann entropy. To make this point even stronger, we will now construct a new entropy of RSF based on the Wehrl entropy and show that for the majority of states it has approximately the same value as the reduced entropy. In other words, despite being based on the quantum von Neumann entropy, the reduced entropy gives the same quantitative results as RSF entropy based on the semi-classical Wehrl entropy.

Making use of the maximum entropy principle, just like in the case of the original reduced entropy, we derive the \emph{reduced Wehrl entropy}
\begin{align} \label{eq:reduced_Wehrl_entropy}
    s_w(\hat{\rho}) \coloneqq \tr \ln (r_\alpha+\mathds{1}_N)+N.
\end{align}
See Appendix \ref{app:entropy} for details. 

Just like the reduced entropy (\ref{eq:reduced_von_Neumann_entropy}) maximizes the von Neumann entropy for a fixed RSF, the reduced Wehrl entropy maximizes the Wehrl entropy for a fixed RSF. We note that it has similar qualitative properties to the original reduced entropy, e.g. it is invariant under the same unitary transformations and is minimized by coherent states.

More importantly, the two entropies can also be linked quantitatively. 
\begin{proposition} \label{th:entropies_comparison}
The following relation between the RSF entropies holds:
\begin{align} \label{eq:reduced_entropies_theorem}
    0 < s_w(\hat{\rho}) - s_v(\hat{\rho}) \leqslant N.
\end{align}
\end{proposition}
\begin{proof}
We begin with the l.h.s. inequality. Rearranging eq. (\ref{eq:reduced_von_Neumann_entropy}) we obtain
\begin{align}
\begin{split}
    s_v(\hat{\rho}) =\;& \tr\left\{r_\alpha\left[\ln(r_\alpha+\mathds{1}_N)-\ln r_\alpha\right]\right\}\\
        & + \tr\ln(r_\alpha+\mathds{1}_N).
\end{split}
\end{align}
By definition of the reduced Wehrl entropy, the second term is equal to $-N+s_w(\hat{\rho})$. In the first term, we apply the eigendecomposition $r_\alpha=\sum_{k=1}^N \lambda_k \ket{k}\bra{k}$, where $\lambda_k\geqslant 0$. Using basic properties of the logarithm, we arrive at
\begin{align}
    s_v(\hat{\rho}) = \sum_{k=1}^N \ln\left(1+1/\lambda_k\right)^{\lambda_k} - N + s_w(\hat{\rho}).
\end{align}      
Clearly, the first term is maximized in the limit $\lambda_k\to\infty$, in which, by definition of the Euler's number, it approaches $N$. Then, the first and second terms cancel, leaving $s_v(\hat{\rho}) < s_w(\hat{\rho})$ as in the l.h.s. inequality. 

To prove the r.h.s. inequality we observe that, since $r_\alpha \geqslant 0$:
\begin{align}
\begin{split}
    s_v(\hat{\rho}) \geqslant \tr\big[&(r_\alpha+\mathds{1}_N)\ln(r_\alpha+\mathds{1}_N)\\
        & - r_\alpha\ln(r_\alpha+\mathds{1}_N)\big] = s_w(\hat{\rho}) - N,
\end{split}
\end{align}
which is equivalent to the r.h.s. inequality.
\end{proof}

Crucially, for states with mean particle number much bigger than the effective number of modes $\tr r = \braket{\hat{n}} \gg N$, the term $N$ is vanishing in comparison to $s_w$, $s_v$. Therefore, it follows from eq. (\ref{eq:reduced_entropies_theorem}) that for most many-particle states, the two reduced entropies are effectively equal. Combining this with our previous analysis of the qualitative aspects of the two entropies, we see that in the RSF formalism, entropic descriptions based on the ``quantum'' von Neumann and on the ``classical'' Wehrl entropy are nearly identical to each other and akin to the Wehrl entropy \footnote{Note that a similar result does not hold for the original entropies, as while the difference $S_W - S_V$ is always positive, there is no known upper bound for it.}. This cements the classicality of the RSF description.

\section{Reduced kinetic equations} 
\label{sec:reduced_kinetic_equations}
Having established the classicality of RSF, we can use the framework to derive and characterize the classical subset of Gaussian evolution. To do this, we will employ the final component of the formalism -- reduced kinetic equations -- which we summarize in this section.

Let us go back to the GKLS equation (\ref{eq:GKLS_diagonal}) and consider its general, non-diagonal form:
\begin{align} \label{eq:GKLS}
    \frac{d}{dt}\hat{\rho}=&
        -i\big[\hat{H},\hat{\rho}\big]
        +\sum_{k,k'}B_{kk'}\left(\hat{J}_k\hat{\rho}\hat{J}_{k'}^\dag
        -\frac{1}{2}\big\{\hat{J}_{k'}^\dag\hat{J}_{k},\hat{\rho}\big\}\right).
\end{align}
Here, $\hat{J}_k$ are the jump operators and $B$ is a non-negative matrix. By diagonalizing the non-Hamiltonian part, the original equation (\ref{eq:GKLS_diagonal}) is obtained.

In the formalism of RSF, it is assumed that to correctly describe the dynamics of a macroscopic field, it is enough to treat it as a set of individual particles subject to spontaneous decay and production, as well as interaction with coherent classical sources and random scattering by the environment.

In such a setting, the following Hamiltonian arises \cite{alicki_reduced_state}:
\begin{align} \label{eq:H}
    \hat{H} = \sum_{k=1}^N \left( \omega_k \hat{a}_k^\dag\hat{a}_k + i\zeta_k\hat{a}_k^\dag-i\zeta_k^*\hat{a}_k\right).
\end{align}
Here, the first term is the base Hamiltonian for bosonic fields, with the positive frequencies $\omega_k$ defining the energy levels of the system. The remaining two terms coincide with the hermitian generator of the displacement operator (\ref{eq:displacement operator}) with argument $\vec{\zeta}$, which is experimentally realized by combining the input state with an $N$-mode coherent state $\ket{\vec{\zeta}}$ in an asymmetric beam-splitter \cite{displacement_operator_experimental_Laiho_2009}. Hence, this part of the Hamiltonian can be interpreted as an interaction with a coherent classical source with complex amplitudes $\zeta_k$.

As for the dissipator, three families of jump operators were considered:
\begin{itemize}
    \item $\hat{J}_k = \hat{a}_k$, that describe spontaneous decay of particles in the field at rates given by matrix $B_{kk'}=\Gamma_{\downarrow}^{k'k}$,
    \item $\hat{J}_k = \hat{a}_k^\dag$, that describe spontaneous production of particles in the field at rates given by matrix $B_{kk'}=\Gamma_{\uparrow}^{kk'}$,
    \item $\hat{J}_k = \hat{U}_k$, with $\hat{U}_k$ being a unitary operator. For a large number of unitary operators, this family describes random scattering \cite{unitary_lindbladians_kossakowski_1972,unitary_lindbladians_Kummerer_1987,unitary_lindbladians_Frigerio_1989,covariance_matrix_beyond_quadratic_linowski_2021} with rates given by $B_{kk'}=\eta_k \delta_{kk'}$, $\eta_k\geqslant 0$, $\sum_{k}\eta_k = 1$ \footnote{In \cite{alicki_reduced_state}, the corresponding summation was replaced by an integral, with the matrix $B$ replaced by a non-negative normalized measure $\mu(du)$ on the unitary group. For notational clarity, here, we keep the more general summation.}.
    
    Note that, while it is not explicitly stated in the original work \cite{alicki_reduced_state}, the results stated there imply that the unitaries $\hat{U}_k$ must each transform the annihilation operators as 
    \begin{align} \label{eq:unitary_assumption}
    \hat{U}_k^\dag\hat{a}_m\hat{U}_k = \sum_{l=1}^N (u_k)_{ml} \hat{a}_l,
    \end{align} 
    where $u_k$ have to be unitary to preserve the canonical commutation relations (\ref{eq:canonical_commutation_relations}).
\end{itemize}

Under the collective influence of all these phenomena, the evolution of the density operator reads \cite{alicki_reduced_state} 
\begin{align} \label{eq:Alicki_evolution_macroscopic}
\begin{split}
    \frac{d}{dt}\hat{\rho}=&
        -i\sum_{k=1}^N\omega_k\big[\hat{a}_k^\dag\hat{a}_k,\hat{\rho}\big]
        +\sum_{k=1}^{N}\big[\zeta_k\hat{a}_k^\dag-\zeta_k^*\hat{a}_k,\hat{\rho}\big]\\
    &+\sum_{k,k'=1}^{N}\Gamma^{k'k}_\downarrow\left(
        \hat{a}_{k}\hat{\rho}\,\hat{a}_{k'}^\dag
        -\frac{1}{2}\big\{\hat{a}_{k'}^\dag\hat{a}_{k},\hat{\rho}\big\}\right)\\
    &+\sum_{k,k'=1}^{N}\Gamma^{kk'}_\uparrow\left(
        \hat{a}_{k}^\dag\hat{\rho}\,\hat{a}_{k'}
        -\frac{1}{2}\big\{\hat{a}_{k'}\hat{a}_{k}^\dag,\hat{\rho}\big\}\right)\\
    &+\sum_k \eta_k\left(\hat{U}_k\hat{\rho}\,\hat{U}_k^\dag-\hat{\rho}\right).
\end{split}
\end{align}
Note that the number of unitaries $\hat{U}_k$ is arbitrary.

Tracing both sides of eq. (\ref{eq:Alicki_evolution_macroscopic}) with $\hat{a}_{l'}^\dag\hat{a}_l$ and $\hat{a}_l$ yields the \emph{reduced kinetic equations} for RSF. As the resulting equations slightly differ from the ones derived originally in \cite{alicki_reduced_state}, where minor errors appear \footnote{In comparison with the original derivation \cite{alicki_reduced_state}, in eq. (\ref{eq:Alicki_evolution_macroscopic}) we define the matrix $\Gamma_\downarrow$ as a transposition of the matrix defined therein, anticipating the final result. After such correction, the reduced kinetic equations (\ref{eq:evolution_standard_RSF}) are nearly identical to the original ones, except the anticommmutator term is divided by two.}, we provide them in full in the following proposition, with proof in Appendix \ref{app:rke}.

\begin{proposition}
The time evolution of RSF is governed by the \emph{reduced kinetic equations}:
\begin{align} \label{eq:evolution_standard_RSF}
\begin{split}
    \frac{d}{dt}{r}=&
        -i\big[h,{r}\big]+\ket{\zeta}\bra{\alpha}+\ket{\alpha}\bra{\zeta}\\
    &+\frac{1}{2}\big\{\gamma_\uparrow-\gamma_\downarrow,{r}\big\}+\gamma_\uparrow\\
    &+\sum_k \eta_k\big(u_k{r}u_k^\dag-{r}\big),\\
    \frac{d}{dt}\ket{\alpha}=&
        -i h\ket{\alpha}
        +\frac{1}{2}\big(\gamma_\uparrow-\gamma_\downarrow\big)\ket{\alpha}+\ket{\zeta}\\
    &+\sum_k \eta_k\big(u_k-1\big)\ket{\alpha}.
\end{split}
\end{align}
Here, 
\begin{align}
    h & \coloneqq \sum_{k=1}^{N}\omega_k\ket{k}\bra{k}, \label{eq:reduced_h} \\
    \ket{\zeta} & \coloneqq \sum_{k=1}^{N}\zeta_k\ket{k}, \label{eq:reduced_zeta} \\
    \gamma_\updownarrow & \coloneqq \sum_{k,k'=1}^{N}\Gamma^{kk'}_{\updownarrow}\ket{k}\bra{k'} \label{eq:reduced_gamma}
\end{align}
are the single-particle counterparts to $\hat{H}$, $\vec{\zeta}$ and ${\Gamma}_\updownarrow$, respectively, while $u_k$ are fixed by eq. (\ref{eq:unitary_assumption}).
\end{proposition}

The assumptions behind the model are best justified by its applicability. In the original work \cite{alicki_reduced_state}, the reduced kinetic equations were successfully used to describe macroscopic fields in thermal environments, as well as polarization optics, notably making an explicit connection between the Mueller and Jones calculi.



\section{Classicality of quantum Gaussian evolution} 
\label{sec:evolution}
Due to their full compatibility with RSF, the reduced kinetic equations necessarily constitute a semi-classical model of evolution. In this section, we use them as a tool for identifying the semi-classical subset of quantum Gaussian evolution, fulfilling the main goal of our work. For clarity, we consider each of the three terms entering the quantum Gaussian evolution equations  (\ref{eq:covariance_evolution}) separately. All proofs are contained in Appendix \ref{app:Gaussian_evolution}.

We begin with the Hamiltonian term.

\begin{proposition} \label{th:classicality_G} 
Let $(V, \ket{\xi})$ denote the symplectic description of a system undergoing Gaussian Hamiltonian evolution
\begin{align} \label{eq:evolution_G}    
    \frac{d}{dt}V = F_G(V), \qquad \frac{d}{dt}\ket{\xi} = f_G(\ket{\xi})
\end{align}
as given by eq. (\ref{eq:F_G}). The evolution can be written as reduced kinetic equations (\ref{eq:evolution_standard_RSF}) and is thus classical with respect to the RSF formalism if and only if
\begin{align} \label{eq:classicality_G}       
    0=\big[J,G\big].
\end{align}
The corresponding reduced kinetic equations are governed by
\begin{align} \label{eq:classicality_G_rke}      
    h = i \mathcal{R}JG\mathcal{R}^\dag
\end{align}
with the remaining terms vanishing.
\end{proposition}

\begin{proof}
See Appendix \ref{app:Gaussian_evolution}.
\end{proof}

To see what the condition (\ref{eq:classicality_G}) means, we make use of the matrix representation of the symplectic form (\ref{symplectic_form_explicit}). Substituting it into (\ref{eq:classicality_G}), we compute that the allowed Hamiltonians consist of $2\times 2$ block matrices of the form
\begin{equation} \label{eq:covariance_matrix_evolution_hamiltonian_restriction_explicit}
\begin{split}           
    G_{kk'}=G_{k'k}^T&=
        a_{kk'}\mathds{1}_2+(1-\delta_{kk'})b_{kk'}J_2,
\end{split}
\end{equation}
where $k$, $k'$ enumerate the blocks and $a_{kk'},b_{kk'}\in\mathbb{R}$. Making use of eq. (\ref{eq:hamiltonian_quadratic}) we check that eq. (\ref{eq:covariance_matrix_evolution_hamiltonian_restriction_explicit}) allows only for particle number-preserving, or \emph{passive} interactions. 

In standard optical implementations, passive transformations correspond to experimental operations with classical analogues, such as beam splitters and phase shifters. According to standard notions of non-classicality, such as non-positivity of the Glauber P representation or the presence of entanglement, the output of passive transformations can be non-classical only if given non-classical input \cite{beam_splitter_classicality_Brunelli_2015,beam_splitter_squeezed_Goldberg_2021}. The remaining \emph{active} transformations, such as squeezing, have no classical analogues. Moreover, they can be a source of quantum advantage, e.g. in metrology \cite{squeezed_metrology, Rudnicki:20}. Such transformations are forbidden by eq. (\ref{eq:covariance_matrix_evolution_hamiltonian_restriction_explicit}).

In short, Gaussian Hamiltonians are classical with respect to the reduced kinetic equations if they correspond to passive transformations.

Let us illustrate this with an example. Among the key ingredients in the resource-based approach to quantum thermodynamics are thermal operations, defined as energy-preserving operations on continuous variable systems coupled to a thermal environment. Due to the prevalence of quadratic Hamiltonians in experimental setups, special emphasis is put on \emph{Gaussian thermal operations} (GTOs), which are thermal operations that preserve the set of Gaussian states.

Recently, a complete characterization of GTOs has been provided in \cite{GTOs}. Here, we focus on a natural subclass of GTOs generated by time-independent, non-degenerate Hamiltonians. Such GTOs are effectively reduced to single-mode transformations \cite{GTOs} of the form
\begin{align} \label{eq:GTO}           
    V(t) = S\left[Q(t)S^{-1} V(0) (S^{-1})^T Q^T(t) + P\right] S^T,
\end{align}
where $S$ is a $2\times 2$ symplectic matrix, $P\coloneqq(1-p)\nu\mathds{1}$ and
\begin{align}   
    Q(t)\coloneqq\sqrt{p}\begin{bmatrix}
        \cos\phi(t) & \sin\phi(t) \\
        -\sin\phi(t) & \cos\phi(t)
        \end{bmatrix}.   
\end{align}
Here, $\nu\coloneqq\coth\beta\omega/2$, $\omega$ is the Hamiltonian eigenvalue associated with the considered mode, $\beta$ is the inverse temperature, while $p\in[0,1]$.

Taking the time derivative of eq. (\ref{eq:GTO}) we get
\begin{align} \label{eq:GTO_correspondence_one-mode}
    G &= \frac{d\phi}{dt} J S S^T J^T,
\end{align}
which, according to Proposition \ref{th:classicality_L}, governs classical evolution if
\begin{align} \label{eq:Gaussian_thermal_conditions}
    0=\big[J,SS^T\big].
\end{align}
One can easily solve this condition explicitly, from which we find that $S$ must be orthogonal in addition to being symplectic, which means that it is passive \cite{GTOs}.

Similar considerations concern Gaussian dissipative evolution based on Lindblad operators linear in mode quadratures.

\begin{proposition} \label{th:classicality_L} 
Let $(V, \ket{\xi})$ denote the symplectic description of a system undergoing Gaussian dissipative evolution stemming from Lindblad operators linear in mode quadratures (\ref{eq:Lindbladians_linear}):
\begin{align} \label{eq:evolution_L}      
    \frac{d}{dt}V = F_L(V), \qquad \frac{d}{dt}\ket{\xi} = f_L(\ket{\xi})
\end{align}
as given by eq. (\ref{eq:F_L}). The evolution can be written as reduced kinetic equations (\ref{eq:evolution_standard_RSF}) and is thus classical with respect to the RSF formalism if and only if
\begin{align} \label{eq:classicality_L_1}       
    0=\big[J,I_C\big].
\end{align}
and
\begin{align} \label{eq:classicality_L_2}       
    \gamma_{\uparrow} &= 
		\mathcal{R}\big(I_C J - J R_C J\big)\mathcal{R}^\dag \geqslant 0, \\
    \gamma_{\downarrow} &= 
		-\mathcal{R}\big(I_C J + J R_C J\big)\mathcal{R}^\dag \geqslant 0.
\end{align}
The corresponding reduced kinetic equations are governed by $\gamma_{\updownarrow}$ as above with the remaining terms vanishing.
\end{proposition}

\begin{proof}
See Appendix \ref{app:Gaussian_evolution}.
\end{proof}

Through eq. (\ref{eq:classicality_L_2}) we can see that the matrix $I_C$ describes the difference between particle creation and annihilation rates, i.e. particle flow: $\mathcal{R}I_CJ\mathcal{R}^\dag = \gamma_{\uparrow}-\gamma_{\downarrow}$. Thus, the first condition (\ref{eq:classicality_L_1}), by full analogy to the one for the Hamiltonian (\ref{eq:classicality_G}) means that the particle flow operator has to be passive. The second condition (\ref{eq:classicality_L_2}) simply requires non-negative particle creation and annihilation rates.

As an example, let us consider \emph{stabilizability} in two-mode entangled Gaussian systems.  In quantum open systems, it is sometimes desirable to counteract the effects of dissipation by using an appropriate Hamiltonian. In the framework of stabilizability, one can check whether this is possible by solving a finite set of conditions rather than checking every Hamiltonian separately \cite{stabilizability_geometric,stabilizability_cv_systems}. 

Recently, stabilizability was used to investigate the robustness of two-mode Gaussian states against three classes of dissipation \cite{stabilizing_entanglement_in_two_mode_Gaussian_states}:
\begin{enumerate}[i.]
    \item local damping: $\hat{L}_k \coloneqq \hat{a}_k$,
    \item global damping: $\hat{L} \coloneqq\left(\hat{a}_1+\hat{a}_2\right)$,
    \item dissipators engineered to preserve two-mode squeezed states:
        \begin{align}
        \begin{split}
            \hat{L}_1 & \coloneqq 
                \cosh\chi\, \hat{a}_1 - \sinh\chi\, \hat{a}_2^\dag, \\
            \hat{L}_2 & \coloneqq
	            \cosh\chi\, \hat{a}_2 - \sinh\chi\, \hat{a}_1^\dag,
        \end{split}
        \end{align}
        where $\chi\geqslant 0$ denotes the amount of squeezing;
\end{enumerate}
It is straightforward to check that while all the dissipators fulfill eq. (\ref{eq:classicality_L_1}), only the first two fulfill the positivity condition (\ref{eq:classicality_L_2}), unless no squeezing is considered in the third model ($\alpha=0$). This, of course, makes sense from the point of classicality, since squeezing is a purely quantum resource, while the Lindblad operators appearing in the first two models merely describe particle loss in the system.

In addition, we remark that in the first and third models, the maximum amount of entanglement was stabilized in the system when using the Hamiltonian
\begin{align}
    \hat{H}_{\textnormal{sq}}\coloneqq
                -i\omega\big(\hat{a}_1\hat{a}_2-\hat{a}_1^\dag\hat{a}_2^\dag\big),
\end{align}
while in the second model the entanglement-maximizing Hamiltonian read
\begin{align}
    \hat{H}=\hat{H}_{\textnormal{cas}}\coloneqq
            -\frac{i\omega}{2}\left[\big(\hat{a}_1+\hat{a}_2\big)^2-
            \big(\hat{a}_1^\dag+\hat{a}_2^\dag\big)^2\right].
\end{align}
As expected, neither Hamiltonian fulfills the classicality condition (\ref{eq:classicality_G}).

Finally, we consider Gaussian dissipative evolution based on unitary Lindblad operators.

\begin{proposition} \label{th:classicality_U} 
Let $(V, \ket{\xi})$ denote the symplectic description of a system undergoing quantum Gaussian dissipative evolution stemming from unitary Lindblad operators (\ref{eq:Lindbladians_unitary}):
\begin{align} \label{eq:evolution_U}      
    \frac{d}{dt}V = F_U(V), \qquad \frac{d}{dt}\ket{\xi} = f_U(\ket{\xi})
\end{align}
as given by eq. (\ref{eq:F_U}). The evolution can be written as reduced kinetic equations (\ref{eq:evolution_standard_RSF}) and is thus classical with respect to the RSF formalism if and only if each $K_j$ fulfills
\begin{align} \label{eq:classicality_U}    
    0=\mathcal{R}K_j\mathcal{R}^T \quad \textnormal{and} 
        \quad \mathcal{R}K_j\mathcal{R}^\dag \textnormal{ is unitary}.
\end{align}
The corresponding reduced kinetic equations are governed by
\begin{align} \label{eq:classicality_U_rke}      
    u_j&=\mathcal{R}K_j\mathcal{R}^\dag, \quad \eta_j = \gamma_j
\end{align}
with the remaining terms vanishing.
\end{proposition}

\begin{proof}
See Appendix \ref{app:Gaussian_evolution}.
\end{proof}

Similarly to previous results, the condition (\ref{eq:classicality_U}) is fulfilled only when the summation is over operations $K_j$, which are orthogonal in addition to being symplectic. From the physical point of view, they also correspond to passive transformations only \cite{GTOs}.

Once again, we illustrate our result with an example. Let us consider the family of two-mode symplectic transformations $K_j =\exp[JS_j]$ generated by
\begin{align}
    S_j = w_j \begin{bmatrix}
     0 & O_j \\
     O_j & 0
    \end{bmatrix},
    \quad
    O_j = \begin{bmatrix}
     \cos\phi_j & \sin\phi_j \\
     \sin\phi_j & -\cos\phi_j
    \end{bmatrix}
\end{align}
where $w_j\geqslant 0$, $\phi_j\in[0,2\pi)$. For $\phi=\pi/2$, $K_j$ is reduced to a transformation used for creation of highly entangled mixtures of Gaussian states in the asymptotic limit in \cite{covariance_matrix_beyond_quadratic_linowski_2021}. We can easily calculate that for all $j$, $\mathcal{R}K_j\mathcal{R}^\dag = \cosh(w_j)\mathds{1}_2$, which is unitary only in the trivial case $w_j=0$. Thus, according to Proposition \ref{th:classicality_U}, the evolution is not classical, as expected given its entangling properties.

\section{Concluding remarks} \label{sec:conclusion}
We studied the classicality of quantum Gaussian evolution, a model of time evolution relevant especially in modern quantum optics and continuous variables-based information processing. We derived an explicit set of conditions under which the evolution is classical. The derived conditions forbid so-called active transformations, such as squeezing, instead allowing only passive transformations, which have an intuitive experimental interpretation in terms of operations treating macroscopic light as a classical wave. Our results were obtained using the recent mesoscopic formalism of the reduced state of the field (RSF), which we redeveloped as a tool for classical description of many-particle bosonic fields.

Based on our findings, we suggest the following directions for further research. To start with, our investigations into the RSF framework could be generalized, e.g., it would be interesting to see if our conjecture regarding lack of entanglement description via RSF can be proved (or disproved). Furthermore, the RSF formalism is based on one- and two-point correlation functions. Can a self-consistent mesoscopic framework based on higher-order correlations be designed? If so, what new insights does it offer, in particular, with respect to classicality?

\begin{acknowledgements}
We would like to thank Pawe{\l} Mazurek, Marcin Karczewski, Gerd Leuchs and Stefano Cusumano for discussion. We acknowledge support by the Foundation for Polish Science (IRAP project, ICTQT, contract no. 2018/MAB/5, co-financed by EU within Smart Growth Operational Programme).
\end{acknowledgements}

\bibliography{report}

\begin{thebibliography}{10}
\providecommand{\url}[1]{\texttt{#1}}
\providecommand{\urlprefix}{URL }
\providecommand{\eprint}[2][]{\url{#2}}

\bibitem{photoelectric_effect_Einstein_1905}
A.~Einstein, \emph{{\"{U}ber einen die Erzeugung und Verwandlung des Lichtes
  betreffenden heuristischen Gesichtspunkt}}, Ann. Phys. \textbf{322}, 132
  (1905).

\bibitem{wave_particle_duality_Bohr_1928}
N.~Bohr, \emph{The quantum postulate and the recent development of atomic
  theory}, Nature \textbf{121}, 580 (1928).

\bibitem{P_representation_Glauber}
R.~J. Glauber, \emph{Coherent and incoherent states of the radiation field},
  Phys. Rev. \textbf{131}, 2766 (1963).

\bibitem{coherent_states_review_Martin-Dussaud_2021}
P.~Martin-Dussaud, \emph{Searching for {C}oherent {S}tates: {F}rom {O}rigins to
  {Q}uantum {G}ravity}, {Quantum} \textbf{5}, 390 (2021).

\bibitem{one_photon_classicality_Markiewicz_2019}
M.~Markiewicz, D.~Kaszlikowski, P.~Kurzy\'{n}ski, A.~W\'{o}jcik, \emph{From
  contextuality of a single photon to realism of an electromagnetic wave}, npj
  Quantum Inf. \textbf{5}, 5 (2019).

\bibitem{one_photon_classicality_Das_2022}
T.~Das, M.~Karczewski, A.~Mandarino, M.~Markiewicz, B.~Wo{\l}oncewicz, et~al.,
  \emph{Remarks about {Bell}-nonclassicality of a single photon}, Phys. Lett. A
  \textbf{435}, 128031 (2022).

\bibitem{quantum_optics_book_Lambropoulos_2007}
P.~Lambropoulos, D.~Petrosyan, \emph{Fundamentals of Quantum Optics and Quantum
  Information}, Springer (2007).

\bibitem{gaussian_optics}
L.~Mandel, E.~Wolf, \emph{Optical Coherence and Quantum Optics}, Cambridge
  University Press (1995).

\bibitem{Gaussian_optics_Olivares_2012}
S.~Olivares, \emph{Quantum optics in the phase space}, Eur. Phys. J. Spec. T.
  \textbf{203}, 3 (2012).

\bibitem{Gaussianity_resource_Albarelli_2018}
F.~Albarelli, M.~G. Genoni, M.~G.~A. Paris, A.~Ferraro, \emph{Resource theory
  of quantum {non-Gaussianity and Wigner} negativity}, Phys. Rev. A
  \textbf{98}, 052350 (2018).

\bibitem{Gaussianity_resource_Takagi_2018}
R.~Takagi, Q.~Zhuang, \emph{Convex resource theory of {non-Gaussianity}}, Phys.
  Rev. A \textbf{97}, 062337 (2018).

\bibitem{QKD_Gaussian_Grosshans_2002}
F.~Grosshans, P.~Grangier, \emph{Continuous variable quantum cryptography using
  coherent states}, Phys. Rev. Lett. \textbf{88}, 057902 (2002).

\bibitem{gaussian_information_1}
X.-B. Wang, T.~Hiroshima, A.~Tomita, M.~Hayashi, \emph{Quantum information with
  {Gaussian} states}, Phys. Rep. \textbf{448}, 1 (2008).

\bibitem{gaussian_information_2}
C.~Weedbrook, S.~Pirandola, R.~Garc\'{\i}a-Patr\'on, N.~J. Cerf, T.~C. Ralph,
  et~al., \emph{Gaussian quantum information}, Rev. Mod. Phys. \textbf{84}, 621
  (2012).

\bibitem{Gaussian_classicality_Wigner_Littlejohn_1986}
R.~G. Littlejohn, \emph{The semiclassical evolution of wave packets}, Phys.
  Rep. \textbf{138}, 193 (1986).

\bibitem{Gaussian_classicality_hybrid_Huber_1988}
D.~Huber, E.~J. Heller, \emph{Hybrid mechanics: {A} combination of classical
  and quantum mechanics}, J. Chem. Phys. \textbf{89}, 4752 (1988).

\bibitem{Gaussian_classicality_hybrid_Huber_1989}
D.~Huber, S.~Ling, D.~G. Imre, E.~J. Heller, \emph{Hybrid mechanics. {II}}, J.
  Chem. Phys. \textbf{90}, 7317 (1989).

\bibitem{Gaussian_classicality_path_integrals_2001}
M.~Baranger, M.~A.~M. de~Aguiar, F.~Keck, H.~J. Korsch, B.~Schellhaa{\ss},
  \emph{Semiclassical approximations in phase space with coherent states}, J.
  Phys. A: Math. Gen. \textbf{34}, 7227 (2001).

\bibitem{Gaussian_classicality_useful_Javanainen_2013}
J.~Javanainen, J.~Ruostekoski, \emph{Emergent classicality in continuous
  quantum measurements}, N. J. Phys. \textbf{15}, 013005 (2013).

\bibitem{Gaussian_classicality_useful_Kong_2016}
X.~Kong, A.~Markmann, V.~S. Batista, \emph{Time-sliced thawed {Gaussian}
  propagation method for simulations of quantum dynamics}, J. Phys. Chem. A
  \textbf{120}, 3260 (2016).

\bibitem{Gaussian_classicality_useful_Graefe_2018}
E.~M. Graefe, B.~Longstaff, T.~Plastow, R.~Schubert, \emph{Lindblad dynamics of
  {Gaussian} states and their superpositions in the semiclassical limit}, J.
  Phys. A: Math. Theor. \textbf{51}, 365203 (2018).

\bibitem{alicki_reduced_state}
R.~Alicki, \emph{Quantum features of macroscopic fields: Entropy and dynamics},
  Entropy \textbf{21}, 705 (2019).

\bibitem{mesoscopic_theory_Grmela_2015}
M.~Grmela, V.~Klika, M.~Pavelka, \emph{Reductions and extensions in mesoscopic
  dynamics}, Phys. Rev. E \textbf{92}, 032111 (2015).

\bibitem{wehrl_entropy}
A.~Wehrl, \emph{On the relation between classical and quantum-mechanical
  entropy}, Reports on Mathematical Physics \textbf{16}, 353  (1979).

\bibitem{wehrl_interpretation}
A.~Or{\l}owski, \emph{Wehrl's entropy and classification of states}, Rep. Math.
  Phys. \textbf{43}, 283  (1999).

\bibitem{two-mode_gaussian_etc_proper_norm}
G.~Adesso, A.~Serafini, F.~Illuminati, \emph{Determination of continuous
  variable entanglement by purity measurements}, Phys. Rev. Lett. \textbf{92},
  087901 (2004).

\bibitem{cv_systems_gaussian_states}
G.~Adesso, S.~Ragy, A.~R. Lee, \emph{Continuous variable quantum information:
  {Gaussian} states and beyond}, Open Syst. Inf. Dyn. \textbf{21}, 1440001
  (2014).

\bibitem{GKS_original}
V.~Gorini, A.~Kossakowski, E.~C.~G. Sudarshan, \emph{Completely positive
  dynamical semigroups of {N}‐level systems}, J. Math. Phys. \textbf{17}, 821
  (1976).

\bibitem{lindblad_original}
G.~Lindblad, \emph{On the generators of quantum dynamical semigroups}, Comm.
  Math. Phys. \textbf{48}, 119 (1976).

\bibitem{lindblad_proof_mathematical}
P.~Pearle, \emph{Simple derivation of the {Lindblad} equation}, Eur. J. Phys.
  \textbf{33}, 805 (2012).

\bibitem{WolfHolevo}
T.~Heinosaari, A.~S. Holevo, M.~M. Wolf, \emph{The semigroup structure of
  {Gaussian} channels}, Quantum Info. Comput. \textbf{10}, 619–635 (2010).

\bibitem{quantum_Gaussianity_Genoni_2013}
M.~G. Genoni, M.~L. Palma, T.~Tufarelli, S.~Olivares, M.~S. Kim, et~al.,
  \emph{Detecting quantum {non-Gaussianity via the Wigner function}}, Phys.
  Rev. A \textbf{87}, 062104 (2013).

\bibitem{quantum_Gaussianity_Hughes_2014}
C.~Hughes, M.~G. Genoni, T.~Tufarelli, M.~G.~A. Paris, M.~S. Kim, \emph{Quantum
  {non-Gaussianity} witnesses in phase space}, Phys. Rev. A \textbf{90}, 013810
  (2014).

\bibitem{quantum_Gaussianity_Walschaers_2021}
M.~Walschaers, \emph{{Non-Gaussian} quantum states and where to find them}, PRX
  Quantum \textbf{2}, 030204 (2021).

\bibitem{covariance_matrix_beyond_quadratic_linowski_2021}
T.~Linowski, A.~Teretenkov, {\L}.~Rudnicki, \emph{Dissipative evolution of
  quantum {Gaussian} states} (2021), arXiv:2105.12644.

\bibitem{unitary_lindbladians_kossakowski_1972}
A.~Kossakowski, \emph{On quantum statistical mechanics of non-{Hamiltonian}
  systems}, Rep. Math. Phys. \textbf{3}, 247 (1972).

\bibitem{unitary_lindbladians_Kummerer_1987}
B.~K\H{u}mmerer, H.~Maassen, \emph{The essentially commutative dilations of
  dynamical semigroups on {$M_n$}}, Comm. Math. Phys. \textbf{109}, 1 (1987).

\bibitem{unitary_lindbladians_Frigerio_1989}
A.~Frigerio, H.~Maassen, \emph{Quantum {Poisson} processes and dilations of
  dynamical semigroups}, Probab. Theory Relat. Fields \textbf{83}, 489 (1989).

\bibitem{using_dissipation_1}
S.~Mancini, H.~M. Wiseman, \emph{Optimal control of entanglement via quantum
  feedback}, Phys. Rev. A \textbf{75}, 012330 (2007).

\bibitem{using_dissipation_2}
K.~Koga, N.~Yamamoto, \emph{Dissipation-induced pure {Gaussian} state}, Phys.
  Rev. A \textbf{85}, 022103 (2012).

\bibitem{stabilizability_cv_systems}
{\L}.~Rudnicki, C.~Gneiting, \emph{Stabilizable {Gaussian} states}, Phys. Rev.
  A \textbf{98}, 032120 (2018).

\bibitem{Note1}
We note the Hamiltonian term and the ``linear'' dissipative term are typically
  combined in a single term given by $A=J(G+I_C)$, so that the evolution reads
  $dV/dt = AV+VA^T +J R_C J^T$. In this work, however, we study the Hamiltonian
  and the dissipative dynamics separately.

\bibitem{Gaussian_solvable_Hu_1992}
B.~L. Hu, J.~P. Paz, Y.~Zhang, \emph{{Quantum Brownian motion in a general
  environment: Exact master equation with nonlocal dissipation and colored
  noise}}, Phys. Rev. D \textbf{45}, 2843 (1992).

\bibitem{Gaussian_solvable_Karrlein_1997}
R.~Karrlein, H.~Grabert, \emph{Exact time evolution and master equations for
  the damped harmonic oscillator}, Phys. Rev. E \textbf{55}, 153 (1997).

\bibitem{Gaussian_solvable_Zhang_2012}
W.-M. Zhang, P.-Y. Lo, H.-N. Xiong, M.~W.-Y. Tu, F.~Nori, \emph{General
  non-{Markovian} dynamics of open quantum systems}, Phys. Rev. Lett.
  \textbf{109}, 170402 (2012).

\bibitem{Gaussian_Green_Wang_2014}
J.-S. Wang, B.~K. Agarwalla, H.~Li, J.~Thingna, \emph{Nonequilibrium
  {Green’s} function method for quantum thermal transport}, Front. Phys.
  \textbf{9}, 673 (2014).

\bibitem{Gaussian_Green_Dhar_2014}
A.~Dhar, K.~Saito, P.~H\"anggi, \emph{Nonequilibrium density-matrix description
  of steady-state quantum transport}, Phys. Rev. E \textbf{85}, 011126 (2012).

\bibitem{quantum_resource_theory_entanglement_HHHH_2009}
R.~Horodecki, P.~Horodecki, M.~Horodecki, K.~Horodecki, \emph{Quantum
  entanglement}, Rev. Mod. Phys. \textbf{81}, 865 (2009).

\bibitem{PPT}
M.~Horodecki, P.~Horodecki, R.~Horodecki, \emph{Separability of mixed states:
  Necessary and sufficient conditions}, Phys. Lett. A \textbf{223}, 1 (1996).

\bibitem{PPT_cv_systems}
R.~Simon, \emph{{Peres}-{Horodecki} separability criterion for continuous
  variable systems}, Phys. Rev. Lett. \textbf{84}, 2726 (2000).

\bibitem{geometry_of_quantum_states}
I.~Bengtsson, K.~\.{Z}yczkowski, \emph{Geometry of Quantum States: {An}
  Introduction to Quantum Entanglement}, Cambridge University Press, 2nd
  edition (2017).

\bibitem{maximum_entropy_principle}
E.~T. Jaynes, \emph{Information theory and statistical mechanics}, Phys. Rev.
  \textbf{106}, 620 (1957).

\bibitem{maximum_entropy_principle_Thurner_2017}
S.~Thurner, B.~Corominas-Murtra, R.~Hanel, \emph{Three faces of entropy for
  complex systems: Information, thermodynamics, and the maximum entropy
  principle}, Phys. Rev. E \textbf{96}, 032124 (2017).

\bibitem{Q_representation}
K.~Husimi, \emph{Some formal properties of the density matrix}, Proc. Phys.
  Math. Soc. Jpn \textbf{22}, 264 (1940).

\bibitem{Wehrl_general_entropy_properties_1978}
A.~Wehrl, \emph{General properties of entropy}, Rev. Mod. Phys. \textbf{50},
  221 (1978).

\bibitem{wehrl_minimum}
E.~H. Lieb, \emph{Proof of an entropy conjecture of {Wehrl}}, Comm. Math. Phys.
  \textbf{62}, 35 (1978).

\bibitem{Note2}
Note that a similar result does not hold for the original entropies, as while
  the difference $S_W - S_V$ is always positive, there is no known upper bound
  for it.

\bibitem{displacement_operator_experimental_Laiho_2009}
K.~Laiho, M.~Avenhaus, K.~N. Cassemiro, C.~Silberhorn, \emph{Direct probing of
  the {Wigner} function by time-multiplexed detection of photon statistics}, N.
  J. Phys. \textbf{11}, 043012 (2009).

\bibitem{Note3}
In \cite {alicki_reduced_state}, the corresponding summation was replaced by an
  integral, with the matrix $B$ replaced by a non-negative normalized measure
  $\mu (du)$ on the unitary group. For notational clarity, here, we keep the
  more general summation.

\bibitem{Note4}
In comparison with the original derivation \cite {alicki_reduced_state}, in eq.
  (\ref {eq:Alicki_evolution_macroscopic}) we define the matrix $\Gamma
  _\downarrow $ as a transposition of the matrix defined therein, anticipating
  the final result. After such correction, the reduced kinetic equations (\ref
  {eq:evolution_standard_RSF}) are nearly identical to the original ones,
  except the anticommmutator term is divided by two.

\bibitem{beam_splitter_classicality_Brunelli_2015}
M.~Brunelli, C.~Benedetti, S.~Olivares, A.~Ferraro, M.~G.~A. Paris,
  \emph{Single- and two-mode quantumness at a beam splitter}, Phys. Rev. A
  \textbf{91}, 062315 (2015).

\bibitem{beam_splitter_squeezed_Goldberg_2021}
A.~Z. Goldberg, K.~Heshami, \emph{How squeezed states both maximize and
  minimize the same notion of quantumness}, Phys. Rev. A \textbf{104}, 032425
  (2021).

\bibitem{squeezed_metrology}
C.~M. Caves, \emph{Quantum-mechanical noise in an interferometer}, Phys. Rev. D
  \textbf{23}, 1693 (1981).

\bibitem{Rudnicki:20}
{\L}.~Rudnicki, L.~L. S\'{a}nchez-Soto, G.~Leuchs, R.~W. Boyd,
  \emph{Fundamental quantum limits in ellipsometry}, Opt. Lett. \textbf{45},
  4607 (2020).

\bibitem{GTOs}
A.~Serafini, M.~Lostaglio, S.~Longden, U.~Shackerley-Bennett, C.-Y. Hsieh,
  et~al., \emph{Gaussian thermal operations and the limits of algorithmic
  cooling}, Physical Review Letters \textbf{124} (2020).

\bibitem{stabilizability_geometric}
S.~Sauer, C.~Gneiting, A.~Buchleitner, \emph{Optimal coherent control to
  counteract dissipation}, Phys. Rev. Lett. \textbf{111}, 030405 (2013).

\bibitem{stabilizing_entanglement_in_two_mode_Gaussian_states}
T.~Linowski, C.~Gneiting, L.~Rudnicki, \emph{Stabilizing entanglement in
  two-mode {Gaussian} states}, Phys. Rev. A \textbf{102}, 042405 (2020).

\bibitem{altland}
A.~Altland, B.~D. Simons, \emph{Condensed Matter Field Theory}, Cambridge
  University Press, 2 edition (2010).

\end{thebibliography}
\bibliographystyle{obib}

\appendix

\section{Derivation of reduction map (\ref{eq:reduced_fields_relation_to_covariance_matrix})} \label{app:reduction_map}
\setcounter{equation}{0}
\renewcommand{\theequation}{\ref{app:reduction_map}\arabic{equation}}
In this appendix, we derive the relation (\ref{eq:reduced_fields_relation_to_covariance_matrix}) between the RSF and covariance matrix pictures (\ref{eq:evolution_standard_RSF}).

Beginning with the definition of the single-particle density matrix (\ref{eq:single_particle_density_matrix_definition}) and the annihilation and creation operators (\ref{eq:creation_annihilation_operators}) we quickly obtain
\begin{align} \label{eq:r_in_terms_of_xi}
\begin{split}
    {r}_{kk'} = \frac{1}{2}\Tr\left[
        \hat{\rho}
        \left(\hat{x}_{k'}\hat{x}_k
        +i\hat{x}_{k'}\hat{p}_k
        -i\hat{p}_{k'}\hat{x}_k
        +\hat{p}_{k'}\hat{p}_k\right)\right].
\end{split}
\end{align}
Looking at eqs (\ref{xi}, \ref{covariance_matrix}), we can see that
\begin{align}
\begin{split}
    {V}_{2k'-1,2k-1} = {V}_{2k-1,2k'-1} &= \Tr\left[
        \hat{\rho}\,\hat{x}_{k'}\hat{x}_{k}\right], \\
    {V}_{2k'-1,2k} = {V}_{2k,2k'-1} &= \frac{1}{2}\Tr\left[
        \hat{\rho}\left(\hat{x}_{k'}\hat{p}_{k}
        +\hat{p}_{k}\hat{x}_{k'}\right)\right], \\
        & = \Tr\left[\hat{\rho}\,\hat{x}_{k'}\hat{p}_{k}\right]
        -\frac{i}{2}\delta_{kk'},\\
    {V}_{2k',2k-1} = {V}_{2k-1,2k'} &= \frac{1}{2}\Tr\left[
        \hat{\rho}\left(\hat{p}_{k'}\hat{x}_{k}
        +\hat{x}_{k}\hat{p}_{k'}\right)\right], \\ 
        & = \Tr\left[\hat{\rho}\,\hat{p}_{k'}\hat{x}_{k}\right]
        +\frac{i}{2}\delta_{kk'},\\
    {V}_{2k',2k} = {V}_{2k,2k'} &= \Tr\left[
        \hat{\rho}\,\hat{p}_{k'}\hat{p}_{k}\right],
\end{split}
\end{align}
where we made use of the canonical commutation relations (\ref{eq:canonical_commutation_relations_xi}). Substituting this into eq. (\ref{eq:r_in_terms_of_xi}), we quickly find that it is equivalent to the relation between $r$ and $V$ in eq. (\ref{eq:reduced_fields_relation_to_covariance_matrix}). The relation between $\ket{\alpha}$ and $\ket{\xi}$ is derived in an analogous fashion.

The Heisenberg uncertainty principle (\ref{eq:Heisenberg_uncertainty_principle_RSF}) is derived by acting on the original eq. (\ref{eq:Heisenberg_uncertainty_principle}) from the left with $\mathcal{R}$ and from the right with $\mathcal{R}^\dag$, and using the easy-to-derive identity $\mathcal{R}J\mathcal{R}^\dag = -i\mathds{1}_{N}$, along with the previously derived eq. (\ref{eq:reduced_fields_relation_to_covariance_matrix}).

\section{Derivation of reduced Wehrl entropy} \label{app:entropy}
\setcounter{equation}{0}
\renewcommand{\theequation}{\ref{app:entropy}\arabic{equation}}
In this appendix, we derive the reduced Wehrl entropy (\ref{eq:reduced_Wehrl_entropy}), defined as the maximum Wehrl entropy among all the states with a fixed RSF.

First, let us observe that RSF has the following representation in terms of the Husimi Q function:
\begin{align} \label{eq:rsf_Husimi}
\begin{split}
    {r}_{kk'} &=\int \frac{d^{2N}\vec{\beta}}{\pi^N}
        \left(\beta_{k}\beta_{k'}^*-\delta_{kk'}\right)Q(\vec{\beta}),\\
    \alpha_k &= \int\frac{d^{2N}\vec{\beta}}{\pi^N}\beta_{k}Q(\vec{\beta}).
\end{split}
\end{align}
The maximum Wehrl entropy among all the states with fixed RSF can then be found by finding the extremum of the following functional with respect to $Q$:
\begin{equation} \label{eq:Wehrl_entropy_variational_problem}
\begin{split}
    S_W[Q]-\lambda & f[Q]-\sum_{k,k'=1}^N\mu_{k'k}g_{kk'}[Q]\\
        &+\sum_{k=1}^Nt_k^*h_k[Q]+\sum_{k=1}^N s_k h_k^*[Q],
\end{split}
\end{equation}
where $S_W$ is the Wehrl entropy (\ref{eq:Wehrl_entropy}) and the three constraints
\begin{equation} \label{eq:Wehrl_entropy_variational_constraints}
\begin{split}
    f[Q]&\coloneqq\int\frac{d^{2N}\vec{\beta}}{\pi^N}Q(\vec{\beta})-1=0,\\
    g_{kk'}[Q]&\coloneqq\int
        \frac{d^{2N}\vec{\beta}}{\pi^N}\left(\beta_{k}\beta_{k'}^*-\delta_{kk'}\right)Q(\vec{\beta})-r_{kk'}=0,\\
    h_k[Q]&\coloneqq \int \frac{d^{2N}\vec{\beta}}{\pi^N}\beta_{k}Q(\vec{\beta})-\alpha_{k}=0
\end{split}
\end{equation}
fix the normalization and the RSF of the state to $(r,\ket{\alpha})$ [cf. eq. (\ref{eq:rsf_Husimi})]. Finally, $\lambda$, $\mu_{k'k}$, $t_k$ and $s_k$ are the Lagrange multipliers. Note that the signs, as well as the notation (e.g. $t_k^*$ instead of $t_k$) in eq. (\ref{eq:Wehrl_entropy_variational_problem}) are arbitrary. Therefore, we made a choice that anticipates the final result best.

The solution to the variatonal problem is given by
\begin{equation} 
\begin{split}
    \tilde{Q}(\vec{\beta})\coloneqq A 
        e^{-\vec{\beta}^\dag\mu\vec{\beta}+\vec{t}^\dag\vec{\beta}+\vec{\beta}^\dag\vec{s}},
\end{split}
\end{equation}
where $A$ is a normalization constant. Substituting the solution into the three constraints (\ref{eq:Wehrl_entropy_variational_constraints}) and making use of the integration formula
\cite{altland}
\begin{align} \label{eq:integration_trick}
    \int \frac{d^{2N}\vec{\beta}}{\pi^N}e^{-\vec{\beta}^\dag{\mu}\vec{\beta}
        +\vec{t}^\dag\vec{\beta}+\vec{\beta}^\dag\vec{s}}
        =\frac{1}{\det\mu}e^{\vec{t}^\dag{\mu}^{-1}\vec{s}},
\end{align}
yields
\begin{equation} 
\begin{split}
    A = \det\mu\, e^{-\vec{t}^\dag\mu^{-1}\vec{s}},\quad
    \mu^{-1}=r_\alpha+\mathds{1}_N,\quad \vec{t}=\vec{s}=\mu\vec{\alpha}
\end{split}
\end{equation}
and in turn
\begin{equation} 
\begin{split}
    \tilde{Q}(\vec{\beta})=\frac{1}{\det(r_\alpha+\mathds{1}_N)}
        e^{-(\vec{\beta}-\vec{\alpha})^\dag(r_\alpha+\mathds{1}_N)^{-1}(\vec{\beta}-\vec{\alpha})}.
\end{split}
\end{equation}
Plugging this into the definition of Wehrl entropy (\ref{eq:Wehrl_entropy}) leads to eq. (\ref{eq:reduced_Wehrl_entropy}).

\section{Derivation of reduced kinetic equations} \label{app:rke}
\setcounter{equation}{0}
\renewcommand{\theequation}{\ref{app:rke}\arabic{equation}}
In this appendix, we derive the reduced kinetic equations (\ref{eq:evolution_standard_RSF}) from the GKLS equation for macroscopic fields (\ref{eq:Alicki_evolution_macroscopic}).

By definition, the single-particle density matrix evolves as
\begin{align}
\begin{split}
    \frac{d}{dt}{r}_{ll'} = \Tr\left( 
        \frac{d}{dt}\hat{\rho}\,\hat{a}_{l'}^\dag \hat{a}_l\right)
         = \sum_{n=1}^5 (\Phi_n)_{ll'}
\end{split}
\end{align}
where [cf. eq. (\ref{eq:Alicki_evolution_macroscopic})]
\begin{align}
    (\Phi_1)_{ll'}  & \coloneqq -i\sum_{k=1}^N\omega_k\Tr\left( 
        \big[\hat{a}_k^\dag\hat{a}_k,\hat{\rho}\big]
        \hat{a}_{l'}^\dag \hat{a}_l\right), \\
    (\Phi_2)_{ll'}  & \coloneqq \sum_{k=1}^N\Tr\left( 
        \big[\zeta_k\hat{a}_k^\dag-\zeta_k^*\hat{a}_k,\hat{\rho}\big]
        \hat{a}_{l'}^\dag \hat{a}_l\right), \\
    (\Phi_3)_{ll'}  & \coloneqq \sum_{k,k'=1}^N\Gamma^{k'k}_\downarrow\Tr\left( 
        \left(\hat{a}_{k}\hat{\rho}\,\hat{a}_{k'}^\dag
        -\frac{1}{2}\big\{\hat{a}_{k'}^\dag\hat{a}_{k},\hat{\rho}\big\}\right)
        \hat{a}_{l'}^\dag \hat{a}_l\right),\\
    (\Phi_4)_{ll'}  & \coloneqq \sum_{k,k'=1}^N\Gamma^{kk'}_\uparrow\Tr\left( 
        \left(\hat{a}_{k}^\dag\hat{\rho}\,\hat{a}_{k'}
        -\frac{1}{2}\big\{\hat{a}_{k'}\hat{a}_{k}^\dag,\hat{\rho}\big\}\right)
        \hat{a}_{l'}^\dag \hat{a}_l\right), \\
    (\Phi_5)_{ll'}  & \coloneqq \sum_k \eta_k \label{eq:Phi_5}
        \Tr\left(\left(\hat{U}_k\hat{\rho}\,\hat{U}_k^\dag-\hat{\rho}\right)
        \hat{a}_{l'}^\dag \hat{a}_l\right).
\end{align}

Let us focus on the first term, $\Phi_1$. From the cyclic property of the trace
\begin{align}
\begin{split}
    (\Phi_1)_{ll'} = -i\sum_{k=1}^N\omega_k\Tr\left( 
        \hat{\rho}\big[\hat{a}_{l'}^\dag \hat{a}_l, \hat{a}_k^\dag\hat{a}_k\big]\right),
\end{split}
\end{align}
The commutator can be easily calculated with the use of the canonical commutation relations (\ref{eq:canonical_commutation_relations}) and the well-known property
\begin{align}
\begin{split}
    \big[\hat{O}_1\hat{O}_2,\hat{O}_3\hat{O}_4\big] 
        = & \: \hat{O}_1\big[\hat{O}_2,\hat{O}_3\big]\hat{O}_4 
        + \big[\hat{O}_1,\hat{O}_3\big]\hat{O}_2\hat{O}_4 \\
        & + \hat{O}_3\hat{O}_1\big[\hat{O}_2,\hat{O}_4\big]
        + \hat{O}_3\big[\hat{O}_1,\hat{O}_4\big]\hat{O}_2, 
\end{split}
\end{align}
valid for arbitrary $\hat{O}_j$.

We obtain
\begin{align}
\begin{split}
    (\Phi_1)_{ll'} = -i(\omega_l-\omega_{l'}) \Tr\left(\hat{\rho}\,\hat{a}_{l'}^\dag \hat{a}_l\right).
\end{split}
\end{align}
Using the definitions (\ref{eq:single_particle_density_matrix_definition}, \ref{eq:reduced_h}), it is easy to show that the above is equivalent to 
\begin{align}
\begin{split}
    (\Phi_1)_{ll'} = -i\sum_{j=1}^N\left(h_{lj} r_{jl'} - r_{lj}h_{jl'}\right)
\end{split}
\end{align}
or simply
\begin{align}
\begin{split}
    \Phi_1 = -i\left[h, r\right].
\end{split}
\end{align}
This shows that the first term on the r.h.s. of eq. (\ref{eq:Alicki_evolution_macroscopic}) transforms into the first term on the r.h.s. of eq. (\ref{eq:evolution_standard_RSF}).

In an analogous way, we can show that
\begin{align}
    \Phi_2 & = \ket{\zeta}\bra{\alpha}+\ket{\alpha}\bra{\zeta},\\
    \Phi_3 & = -\frac{1}{2}\big\{\gamma_\downarrow,{r}\big\},\\
    \Phi_4 & = \frac{1}{2}\big\{\gamma_\uparrow,{r}\big\}+\gamma_\uparrow.
\end{align}
As for $\Phi_5$, due to normalization of $\eta_k$ to one, the second term under the trace in eq.  (\ref{eq:Phi_5}) gives rise to simply $r_{ll'}$. The first term, on the other hand, can be rewritten using the cyclic property of the trace and the fact that $\hat{U}_k\hat{U}_k^\dag = \hat{\mathds{1}}$:
\begin{align}
\begin{split}
    (\Phi_5)_{ll'}  & = \sum_k \eta_k
        \Tr\left(\hat{\rho}\,
        \left(\hat{U}_k^\dag\hat{a}_{l'}^\dag \hat{U}_k\right) 
        \left(\hat{U}_k^\dag \hat{a}_l \hat{U}_k\right) \right) - r_{ll'}.
\end{split}
\end{align}
Making use of the assumption (\ref{eq:unitary_assumption}), we quickly find that
\begin{align}
\begin{split}
    \Phi_5  & = \sum_k \eta_k\big(u_k{r}u_k^\dag-{r}\big).
\end{split}
\end{align}
This finishes the derivation of the reduced kinetic equation (\ref{eq:evolution_standard_RSF}) for $r$. The corresponding equation for $\ket{\alpha}$ is derived in the same way.

\section{Derivation of classical Gaussian evolution} \label{app:Gaussian_evolution}
\setcounter{equation}{0}
\renewcommand{\theequation}{\ref{app:Gaussian_evolution}\arabic{equation}}
In this appendix, we prove Propositions \ref{th:classicality_G}, \ref{th:classicality_L} and \ref{th:classicality_U}, i.e. we derive the conditions under which Gaussian evolution is equivalent to the reduced kinetic equations. 

To this end, it will be useful to define an auxiliary field, which we call \emph{conjugate RSF}:
\begin{align} \label{eq:conjugate_RSF}
\begin{split}
    {c}&\coloneqq\sum_{k,k'=1}^{N}
        \Tr\big[\hat{\rho}\,\hat{a}_{k'}\hat{a}_k\big]
        \ket{k}\bra{k'},\\
    \ket{\alpha^*}&\coloneqq\sum_{k=1}^{N}
        \Tr\big[\hat{\rho}\,\hat{a}_k^\dag\big]
        \ket{k}.
\end{split}
\end{align}
Mirroring the derivation of the relation (\ref{eq:reduced_fields_relation_to_covariance_matrix}) between RSF and the symplectic picture, one can show that
\begin{align} \label{eq:reduced_fields_relation_to_covariance_matrix_conjugate}
\begin{split}
    {c}&=\mathcal{R}V\mathcal{R}^T,
        \qquad\ket{\alpha^*}=\mathcal{R}^*\ket{\xi}.
\end{split}
\end{align}
We will also make heavy use of the following property of the reduction matrix:
\begin{align} 
    \mathcal{R}^\dag \mathcal{R} = \frac{1}{2}\left({\mathds{1}}+iJ\right).
\end{align}
Notably,
\begin{align} \label{eq:reduction_matrix_identity}
    \mathcal{R}^\dag \mathcal{R}+\mathcal{R}^T \mathcal{R}^*=\mathds{1}.
\end{align}

\subsection{Proof of Proposition \ref{th:classicality_G}} 
We begin with the Hamiltonian evolution (\ref{eq:evolution_G}). Making extensive use of the identity (\ref{eq:reduction_matrix_identity}), along with relations (\ref{eq:reduced_fields_relation_to_covariance_matrix}, \ref{eq:reduced_fields_relation_to_covariance_matrix_conjugate}), we obtain the corresponding evolution equations for RSF:
\begin{align}
\begin{split}
    \frac{d}{dt}{r}=&\:y_G{r}-{r}y_G^\dag+z_G{c}^\dag+{c}z_G^\dag
        +\frac{1}{2}\left(y_G-y_G^\dag\right),\\
    \frac{d}{dt}\ket{\alpha}=&\:y_G\ket{\alpha}+z_G\ket{\alpha^*},
\end{split}
\end{align}
where
\begin{align}
\begin{split}
    y_G&\coloneqq \mathcal{R}JG\mathcal{R}^\dag, \qquad z_G\coloneqq \mathcal{R}JG\mathcal{R}^T.
\end{split}
\end{align}
Unlike the reduced kinetic equations, this evolution equation for RSF couples it to the conjugate field. Therefore, if the two equations are to coincide for arbitrary input states, the $c$-dependent terms must vanish. This implies $z_G=0$ and in turn $0=\mathcal{R}^\dag z_G \mathcal{R}^*$, which is equivalent to the condition (\ref{eq:classicality_G}), as we intended to show.

Under this condition $y_G$ is hermitian, and hence the final equations read
\begin{align}
\begin{split}
    \frac{d}{dt}{r}=&\:[y_G,{r}] \qquad
    \frac{d}{dt}\ket{\alpha}=\:y_G\ket{\alpha}.
\end{split}
\end{align}
Clearly, they have the form of the reduced kinetic equations (\ref{eq:evolution_standard_RSF}) with eq. (\ref{eq:classicality_G_rke}) at the input.

\subsection{Proof of Proposition \ref{th:classicality_L}} 
In the case of the dissipative evolution stemming from linear Lindblad operators (\ref{eq:evolution_L}), using eqs (\ref{eq:reduced_fields_relation_to_covariance_matrix}, \ref{eq:reduced_fields_relation_to_covariance_matrix_conjugate},\ref{eq:reduction_matrix_identity}) as previously yields
\begin{align}
\begin{split}
    \frac{d}{dt}{r}=&\:y_L{r}+{r}y_L^\dag+z_L{c}^\dag+{c}z_L^\dag
        +\frac{1}{2}\left(y_L+y_L^\dag\right),\\
    \frac{d}{dt}\ket{\alpha}=&\:y_L\ket{\alpha}+z_L\ket{\alpha^*},
\end{split}
\end{align}
where
\begin{align}
\begin{split}
    y_L&\coloneqq \mathcal{R}JI_C\mathcal{R}^\dag, 
    \quad z_L\coloneqq \mathcal{R}JI_C\mathcal{R}^T, 
    \quad w \coloneqq \mathcal{R}JR_CJ^T\mathcal{R}^\dag.
\end{split}
\end{align}
Once again, we must require the equation to be $c$-independent. This implies $z_L=0$ and in turn $0=\mathcal{R}^\dag z_L \mathcal{R}^*$, which is the same as the condition (\ref{eq:classicality_L_1}) that we wanted to derive.

Under this condition $y_L$ is hermitian, and hence the final equations read
\begin{align}
\begin{split}
    \frac{d}{dt}{r}=&\:\{y_L,{r}\} + y_L + w\qquad
    \frac{d}{dt}\ket{\alpha}=\:y_L\ket{\alpha}.
\end{split}
\end{align}
It is not difficult to show that these equations have the form of the reduced kinetic equations (\ref{eq:evolution_standard_RSF}) with eq. (\ref{eq:classicality_L_2}) at the input. Note that in order for this identification to have a physical meaning, the particle creation and annihilation rates have to be non-negative.

\subsection{Proof of Proposition \ref{th:classicality_U}} 
Finally, we consider the dissipative evolution stemming from unitary Lindblad operators (\ref{eq:evolution_U}). Once again making use of eqs (\ref{eq:reduced_fields_relation_to_covariance_matrix}, \ref{eq:reduced_fields_relation_to_covariance_matrix_conjugate}, \ref{eq:reduction_matrix_identity}) we obtain
\begin{align}
\begin{split}
    \frac{d}{dt}{r}=&\sum_j \gamma_j \bigg[q_j r q_j^\dag 
        + s_j r^T s_j^\dag - r + q_j c s_j^\dag + s_j c^* q_j^\dag, \\
        &\qquad\qquad\qquad\qquad+\frac{1}{2}
        \left(q_j q_j^\dag + s_j s_j^\dag-\mathds{1}\right)\bigg],\\
    \frac{d}{dt}\ket{\alpha}=&\sum_j \gamma_j \left[
        (q_j-\mathds{1})\ket{\alpha}+s_j\ket{\alpha^*}\right],
\end{split}
\end{align}
where
\begin{align}
\begin{split}
    q_j = \mathcal{R} K_j \mathcal{R}^\dag, \qquad s_j = \mathcal{R} K_j \mathcal{R}^T.
\end{split}
\end{align}
Calculating analogously as in the previous cases, we find that the equation is $c$-independent if for all $j$
\begin{align}     
    0=\mathcal{R}K_j\mathcal{R}^T.
\end{align}
To get a correspondence with the reduced kinetic equations, we must additionally require all $q_j$ to be unitary. The two conditions are collectively captured by eq. (\ref{eq:classicality_U}), finishing the proof.

The final equations read
\begin{align}
\begin{split}
    \frac{d}{dt}{r}=&\sum_j \gamma_j \left(q_jrq_j^\dag-r\right),\\
    \frac{d}{dt}\ket{\alpha}=&\sum_j \gamma_j\left(q_j\ket{\alpha}-\ket{\alpha}\right),
\end{split}
\end{align}
which have the form of the reduced kinetic equations (\ref{eq:evolution_standard_RSF}) with eq. (\ref{eq:classicality_U_rke}) at the input.

\end{document}